\documentclass{aa}  

\bibpunct{(}{)}{;}{a}{}{,} 
\usepackage{graphicx}
\usepackage[varg]{txfonts}
\usepackage{comment}
\usepackage{gensymb}

\usepackage{hyperref}
\hypersetup{%
  colorlinks=true,
  linkcolor=blue,
  urlcolor=blue,
  citecolor=blue,
  bookmarksnumbered=true,
  bookmarksopenlevel=1,
  }

\begin{document} 

\title{Spectra of standing kink waves in loops and the effects of the lower solar atmosphere}

\author{%
{Konstantinos Karampelas}\inst{\ref{aff:CmPA}} \orcid{0000-0001-5507-1891}
\and {Daye Lim}\inst{\ref{aff:CmPA},\ref{aff:ROB}}\orcid{0000-0001-9914-9080}
\and {Tom Van Doorsselaere}\inst{\ref{aff:CmPA}} \orcid{0000-0001-9628-4113}
\and {Yuhang Gao}\inst{\ref{aff:PU},\ref{aff:CmPA}}\orcid{0000-0002-6641-8034}
}

\institute{%
\label{aff:CmPA}{Centre for mathematical Plasma Astrophysics, Department of Mathematics, KU Leuven, Celestijnenlaan 200B, 3001 Leuven, Belgium.}\\ \email{kostas.karampelas@kuleuven.be}
\and
\label{aff:ROB}{Solar-Terrestrial Centre of Excellence – SIDC, Royal Observatory of Belgium, Ringlaan -3- Av. Circulaire, 1180 Brussels, Belgium}
\and
\label{aff:PU}{School of Earth and Space Sciences, Peking University, Beijing, 100871, People’s Republic of China}
}

\date{Received \today; Accepted \today}

\abstract
{Understanding the effects of the lower solar atmosphere on the spectrum of standing kink oscillations of coronal loops, in both  the decaying and decayless regime, is essential for developing more advanced tools for coronal seismology.}
{We aim to reveal the effects of the chromosphere on the spatial profiles and frequencies of the standing kink modes, create synthetic emission maps to compare with observations, and study the results using spatial and temporal coronal seismology techniques.}
{We excited transverse oscillations in a 3D straight flux tube using (a) a broadband footpoint driver, (b) a sinusoidal velocity pulse, and (c) an off-centre Gaussian velocity pulse, using the PLUTO code. The flux tube is gravitationally stratified, with footpoints embedded in chromospheric plasma. Using the FoMo code, we created synthetic observations of our data in the Fe IX 17.1 nm line and calculated the spectra with the Automatic Northumbria University Wave Tracking code. We also numerically solved the generalised eigenvalue system for the 1D wave equation to determine the effects of the stratification on the kink modes of our system.}
{The synthetic observations of the loops perturbed by the velocity pulses show a single dominant mode that our 1D analysis reveals to be the third harmonic of the system. For the broadband driver, the synthetic emission shows multiple frequency bands, associated with both the loop and the driver. Finally, using seismological techniques, we highlight the possibility of misidentifying the observed third, sixth, and ninth harmonics with the first, second, and third harmonics of the coronal part of the loop. Unless more advanced techniques of spatial seismology are used with many data points from observations along the entire loop length, this misidentification can result in overestimating the mean magnetic field by a factor equal to the period ratio of the fundamental over the third harmonic.}
{For longer coronal loops it is easy to misidentify the detected standing kink modes for lower-order modes of the system, which can have important seismological implications. To prevent these errors and properly constrain the value of the estimated mean magnetic field, additional observations of the loops footpoints using transition region and chromospheric lines are necessary.}

\keywords{magnetohydrodynamics - solar coronal seismology - solar coronal waves - magnetohydrodynamical simulations}

\titlerunning{Oscillation spectra of stratified loops}
\authorrunning{Karampelas et al.}

\maketitle
\nolinenumbers
\section{Introduction} \label{sec:introduction}

Measuring quantities such as the mean magnetic field or the density in the solar atmosphere is a task that requires a combination of observations, theory, and numerical modelling to extract information about the local plasma parameters. Coronal seismology \citep[e.g.][]{Roberts1984ApJ...279..857R,Nakariakov2005LRSP} uses  observed waves in the solar atmosphere, which are understood to be magnetohydrodynamic (MHD) waves in inhomogeneous plasma \citep{edwin1983wave}, to perform such calculations. Space-based telescopes such as the Transition and Coronal Explorer \citep[TRACE;][]{Handy1999SoPh..187..229H}, the Solar Dynamics Observatory/Atmospheric Imaging Assembly \citep[SDO/AIA;][]{Lemen2012SoPh..275...17L}, and the Solar Orbiter/Extreme Ultraviolet Imager \citep[EUI;][]{Rochus2020A&A...642A...8R} have revealed an omnipresence of transverse oscillations that are understood to be kink waves \citep{tvd2008detection} and which have been extensively used for plasma diagnostics. Observations of kink waves have revealed both propagating motions in open and closed field regions \citep[for a review, see][]{Morton2023RvMPP...7...17M} and standing oscillations in structures such as coronal loops \citep[see][for a review]{NakariakovEtAl2021}.

For the case of standing kink oscillations in loops, two different regimes have been discovered. The first is the decaying regime first reported in \citet{aschwanden1999} and \citet{nakariakov1999}. It is characterised by quickly decaying oscillation amplitudes, with the oscillations being excited by external energetic phenomena \citep[e.g.][]{Nechaeva2019ApJS}, coronal rain \citep{Verwichte2017A&A...598A..57V,VerwichteKohutova2017A&A...601L...2V,KohutovaVerwichte2017A&A...606A.120K}, and nanojets \citep{Sukarmadji2024ApJ...961L..17S}. In addition, the coexistence of multiple harmonics in decaying oscillations of loops has been demonstrated \citep[e.g.][]{DeMoortel2007ApJ...664.1210D,Duckenfield2019A&A...632A..64D}. The second regime consists of decayless oscillations \citep[e.g.][]{wang2012,tian2012}, which are characterised by a near-constant oscillation amplitude over many periods \citep{nistico2013}. These standing oscillations are observed in the solar atmosphere in loops of different scales \citep[e.g.][]{wang2012,nistico2013,anfinogentov2013,anfinogentov2015,anfinogentov2019ApJ,ZhongSihui2022MNRAS.513.1834Z,ZhongSihui2022MNRAS.516.5989Z,LiandLong2023ApJ...944....8L,ZhongLongLoop2023NatSR..1312963Z}. However, their interpretation as standing waves is still a matter of debate for most observations in shorter transition region loops and coronal loops \citep[][]{GaoYuhang2022ApJ...930...55G,Petrova2023ApJ...946...36P,GaoYuhang2024A&A...681L...4G,ShrivastavArpitKumar2024A&A...685A..36S}. It was recently shown in  \citet{Lim2024A&A...690L...8L} that under-sampling due to low cadence can lead to an overestimation of the oscillation periods in shorter loops, giving rise to the observed lack of a correlation between the periods and loop lengths for shorter loops.

Simulations of decayless oscillations have hinted at an energy content sufficient to counterbalance the radiative losses in the quiet Sun corona \citep[e.g.][]{karampelas2019amp,mijie2021ApJL,DeMoortel2022ApJ...941...85D}. Recent meta-analysis studies of observations \citep[e.g.][]{Lim2023ApJ...952L..15L,Lim2024A&A...689A..16L} have found that the high-frequency regime of decayless oscillations can heat the quiet Sun corona, but cannot support the wave heating in active regions unless unresolved oscillations with frequencies up to $0.17$ Hz are considered. Decayless standing waves are also reported to exhibit linear polarisation \citep[see e.g.][for a combined observation with SDO/AIA and the High Resolution Imager HRI$_\mathrm{EUV}$ of Solar Orbiter/EUI]{ZhongPolarisation2023NatCo..14.5298Z}, while the coexistence of multiple harmonics (first and second) was also reported in \citet{duckenfield2018ApJ}. Equally important is that decayless oscillations still have an unidentified excitation mechanism, with different hypotheses proposing it to be broadband footpoint drivers \citep{afanasyev2020decayless,Ruderman2021MNRAS.501.3017R,Ruderman2021SoPh..296..124R,Howson2023Physi...5..140H,Karampelas2024A&A...681L...6K}, p-modes \citep{Skirvin2023ApJ...949...38S,Gao2023ApJ...955...73G}, plasma flows along the loop \citep{KohutovaVerwichte2018A&A...613L...3K}, and external flows \citep{nakariakov2009,nakariakov2016,karampelas2020ApJ,karampelas2021ApJ...908L...7K}. Additionally, \citet{antolin2016} suggest that apparent observations of decayless oscillations are actually observations of decaying oscillations muddled by line-of-sight effects.

As stated earlier, transverse kink oscillations have been used extensively as tools in coronal seismology. For example, by measuring the period of the fundamental kink mode and the loop length, we can calculate the kink speed of the loop and then estimate the average Alfv\'{e}n speed and the average magnetic field using common seismology techniques \citep[e.g.][]{edwin1983wave,Roberts1984ApJ...279..857R,AschwandenSchrijver2011ApJ...736..102A,anfinogentov2019ApJ,ZhongLongLoop2023NatSR..1312963Z, GaoYuhang2024A&A...681L...4G}. Longitudinal density stratification and the expansion of the magnetic field can also lead to the modification of the spatial profile and the respective frequency of the fundamental kink mode and its overtones (e.g. \citealt{andries2005}, see also \citealt{Andries2009SSRv..149....3A} for a review). For example, \citet{Verth2007A&A...475..341V} showed that the antinode of the second harmonic is shifted towards the loop footpoints as the density scale height decreases, for a flux tube with a straight magnetic field \citep[see also][]{ErdelyiVerth2007A&A...462..743E}. If multiple kink mode harmonics are present in the same oscillating loop, then their ratios can be used to determine the density scale height, assuming again that there is no variation in the magnetic field \citep{Andries2005ApJ...624L..57A,Safari2007}. The effects of the lower solar atmosphere and the transition region have started to be incorporated in 3D simulations of open flux tubes and coronal loops in studies of wave propagation \citep{pelouze2023A&A...672A.105P}, excitation of decayless oscillations \citep[e.g.][]{Gao2023ApJ...955...73G,Karampelas2024A&A...681L...6K}, and wave energy dissipation \citep[e.g.][]{mingzhe2023ApJ...949L...1G,Karampelas2024A&A...688A..80K}. \citet{HowsonBrue2023MNRAS.526..499H} have numerically shown from the 1D Sturm-Liouville problem describing an oscillating loop \citep{DymovaRuderman2005SoPh..229...79D} that the existence of the transition region and the chromosphere will greatly modify the profile of the harmonics of standing Alfv\'{e}n and kink waves. It was also suggested that this deformation could be potentially misleading when trying to identify the oscillation modes in observations of loop systems.

We studied the effects of the lower solar atmosphere on the spatial profiles and frequencies of standing kink modes of coronal loops driven by noisy footpoint motions. Our aim is to identify the individual modes in our simulation results and create synthetic emission maps that can be compared with observations. We also applied the techniques of coronal seismology to understand the importance of the lower atmosphere when using kink oscillations as plasma diagnostic tools for quantities such as the kink speed and mean magnetic field. In Sect. \ref{sec:setup} we describe the initial conditions of our 3D coronal loop model and how we excite the oscillations of interest. Section \ref{sec:results} describes the results of the simulations, the creation and analysis of our synthetic observations, the equivalent 1D eigenvalue problem, and the use of seismological tools. Finally, Sect. \ref{sec:discussions} includes a discussion of the most important points of this study.

\section{Numerical setup} \label{sec:setup}

\begin{figure*}
    \centering
    \resizebox{\hsize}{!}{
    \includegraphics[trim={3.cm 0.cm 0.cm 0.cm},clip,scale=0.55]{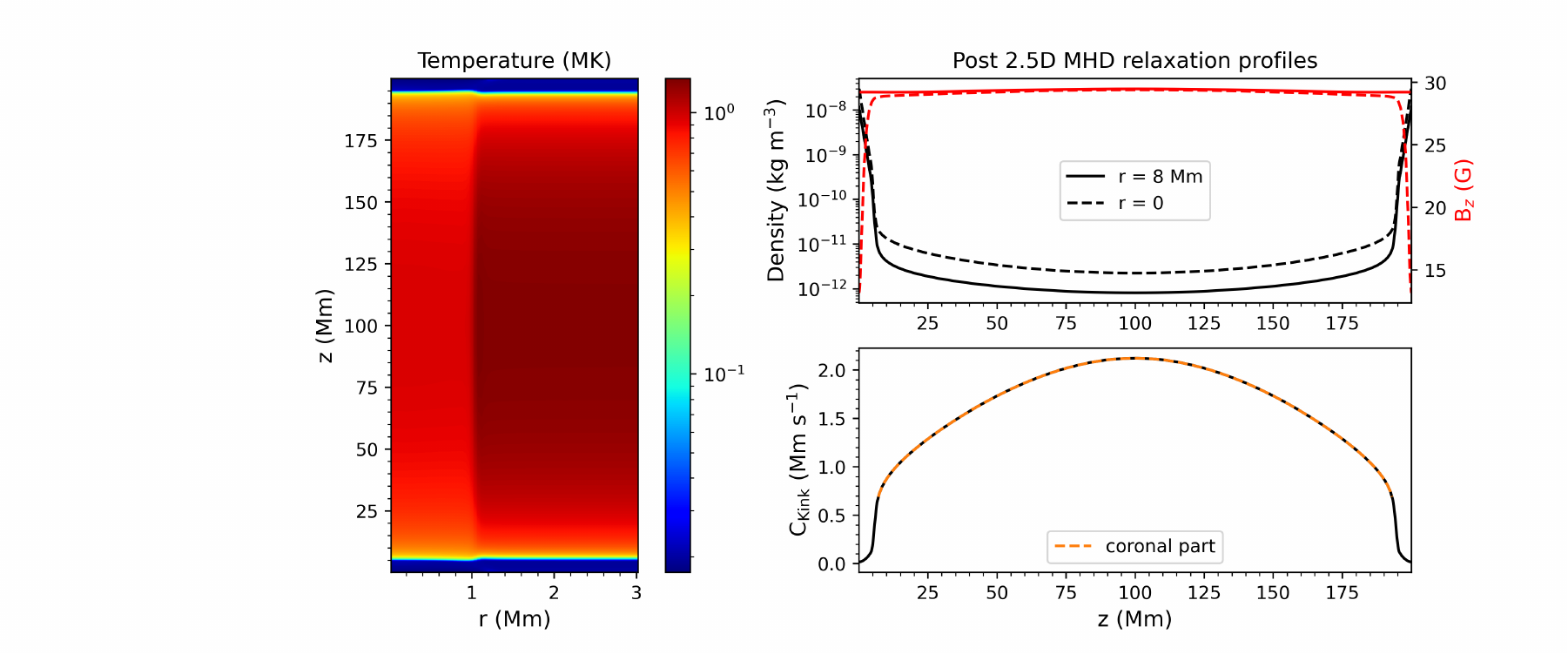}}
    \caption{Post 2.5D MHD relaxation profiles for the temperature, density, $B_z$ magnetic field, and kink speed of our stratified flux tube. Left: 2D temperature profile for a section of our model up to $r=3$\, Mm. Top right: 1D profiles along the $z$ direction of the density (black lines) and $B_z$ magnetic field (red lines) at $r=0$ (solid lines) and at $r=8$\,Mm (dashed lines). Bottom right: Profile of the kink speed along the $z$-axis, with the coronal part ($z\in \left[ 7, 193 \right]$\,Mm) highlighted (dashed orange line).}    \label{fig:inicon}
\end{figure*}

\begin{figure*}
    \centering
    \resizebox{\hsize}{!}{
    \includegraphics[trim={-1.cm 0.cm -1.cm 0.cm},clip,scale=0.55]{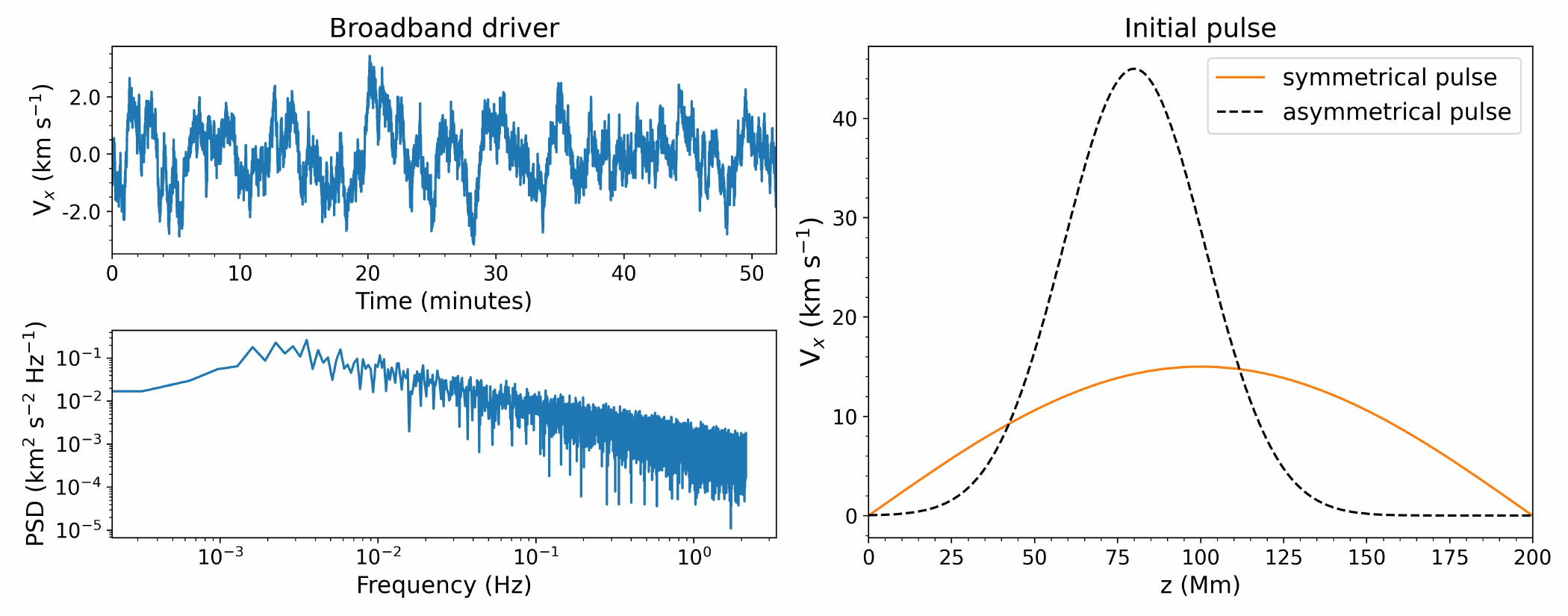}}
    \caption{Profiles of the velocity driver and pulses used in the three 3D simulations. Left panels: Temporal profile of the $v_x$ driver at $z=0$ for the first simulation (top) and its corresponding power spectrum (bottom). Right panel: Two initial velocity conditions used in the two simulations without the footpoint driver.}
    \label{fig:driver}
\end{figure*}

\begin{figure*}
    \centering
    \resizebox{\hsize}{!}{
    \includegraphics[trim={0.cm .3cm 0.cm .8cm},clip,scale=0.55]{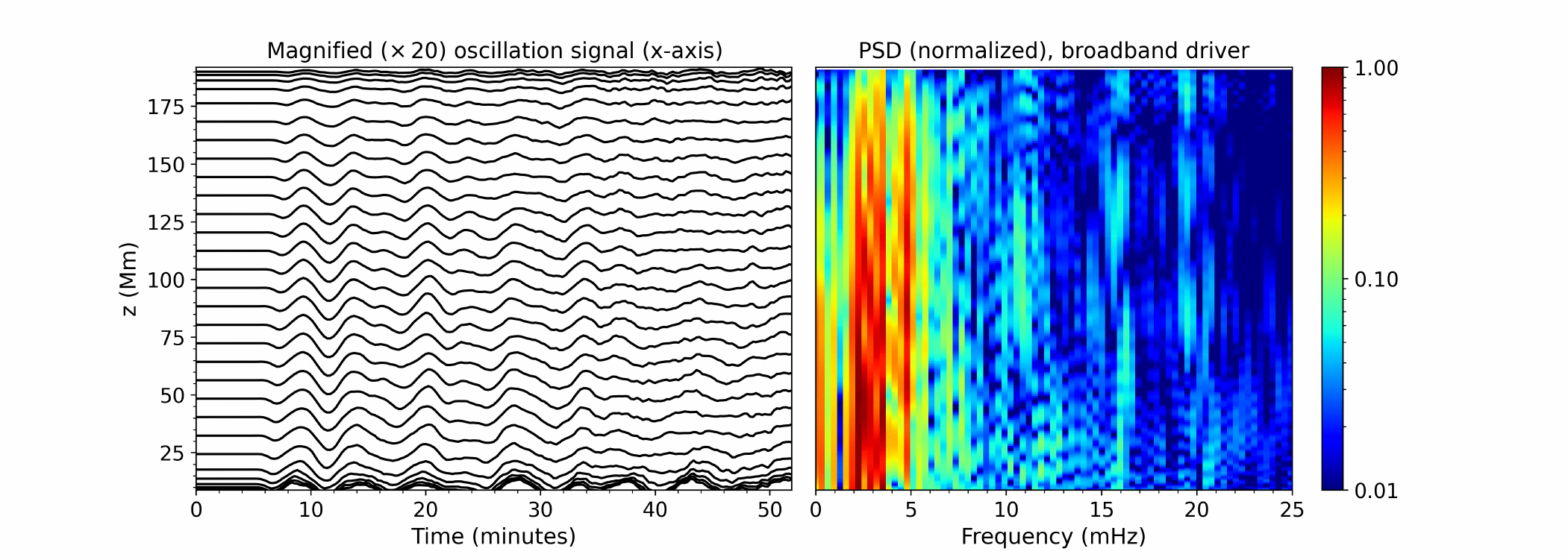}}
    \resizebox{\hsize}{!}{
    \includegraphics[trim={0.cm .3cm 0.cm .8cm},clip,scale=0.55]{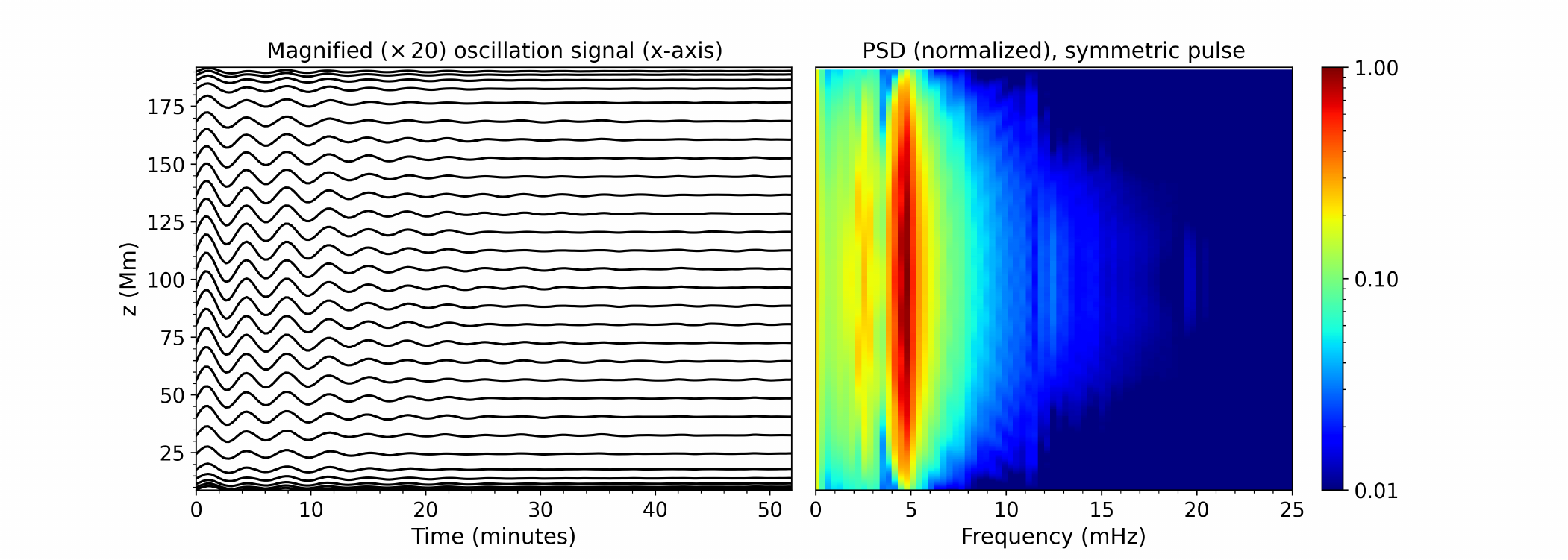}}
    \resizebox{\hsize}{!}{
    \includegraphics[trim={0.cm 0.cm 0.cm .8cm},clip,scale=0.55]{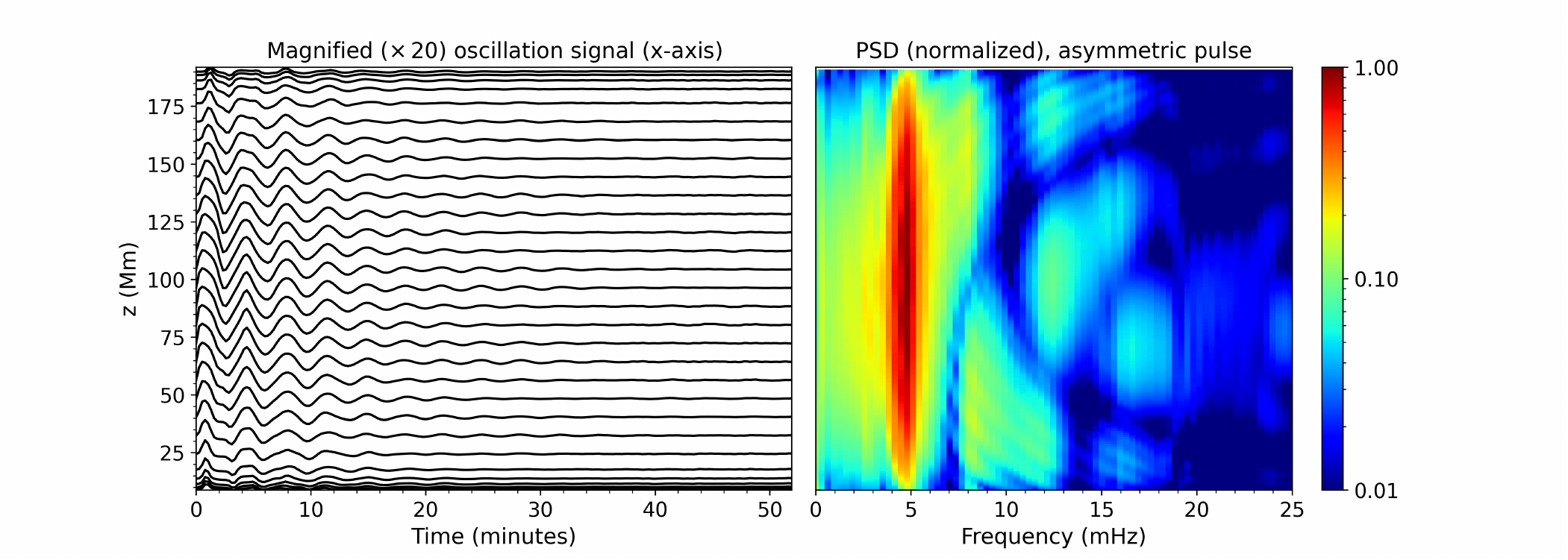}}
    \caption{Centre of mass displacement (left panels) and corresponding power spectra density profiles (right panels) for the three simulations of our 3D loop. The time-distance maps of the displacement along the $x$-direction show the magnified ($\times 20$) signal projected along the $z$-axis for visualisation purposes. Only the coronal part of the loop is depicted. Shown are the simulations with the driver (top), with the symmetric initial pulse (middle), and with the asymmetric pulse (bottom).}
    \label{fig:cmspectra}
\end{figure*}

\begin{figure*}
    \centering
    \resizebox{\hsize}{!}{
    \includegraphics[trim={0.cm 4.cm 0.cm 0.cm},clip,scale=0.5]{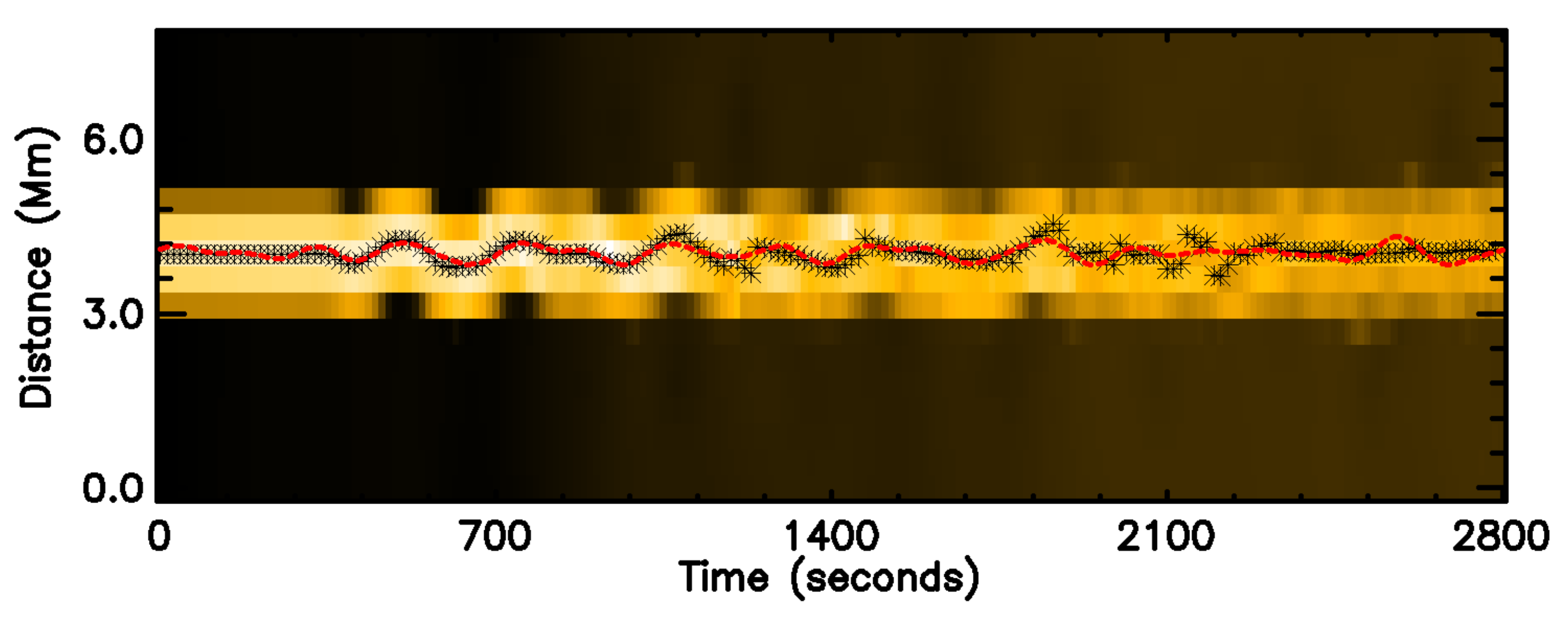}
    \includegraphics[trim={.5cm 4.cm 0.cm 0.cm},clip,scale=0.5]{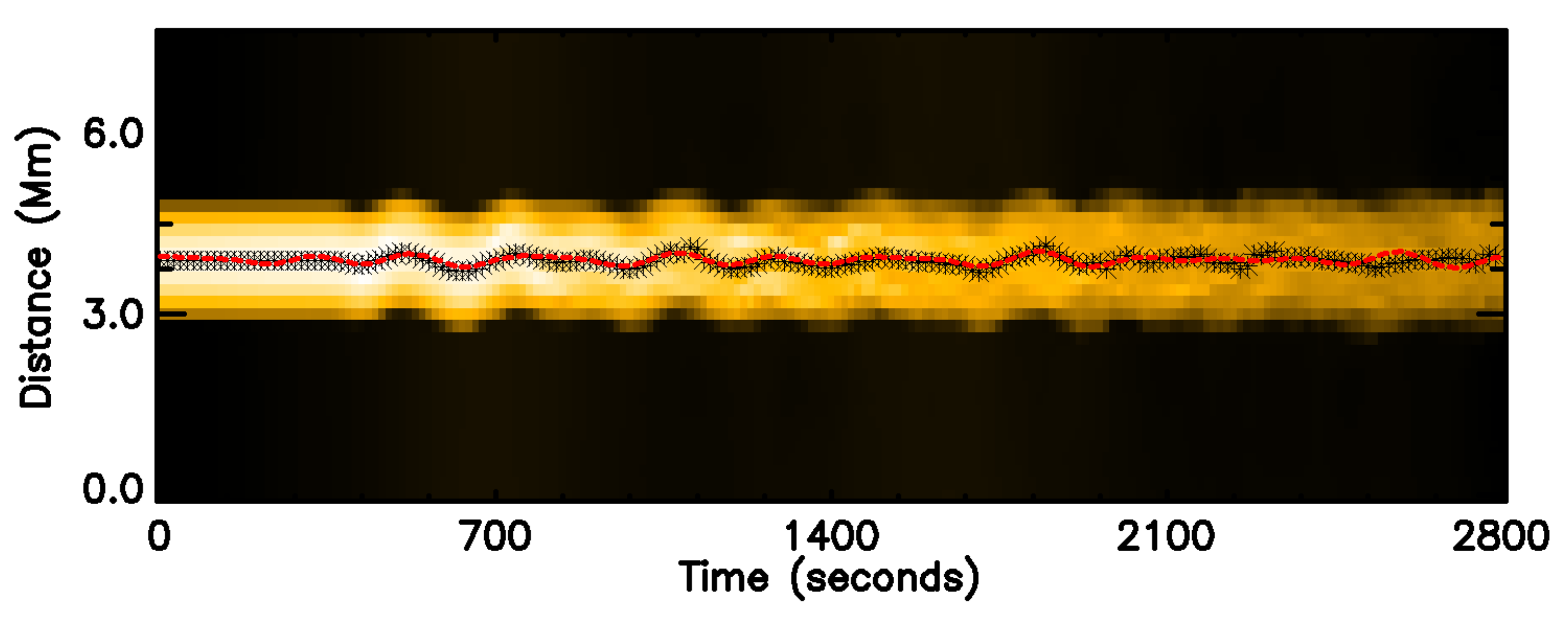}
    \includegraphics[trim={.5cm 4.cm 0.cm 0.cm},clip,scale=0.5]{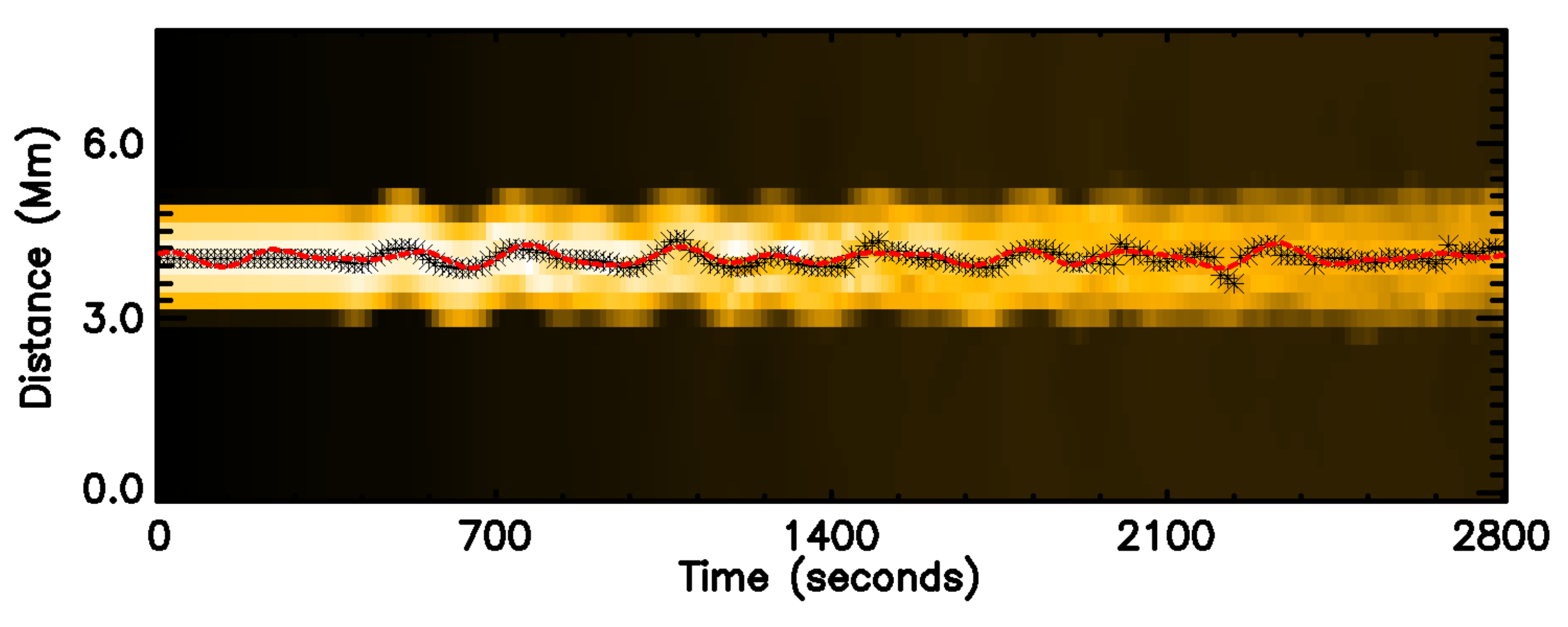}
    }
    \resizebox{\hsize}{!}{
    \includegraphics[trim={0.cm 4.cm 0.cm 0.cm},clip,scale=0.5]{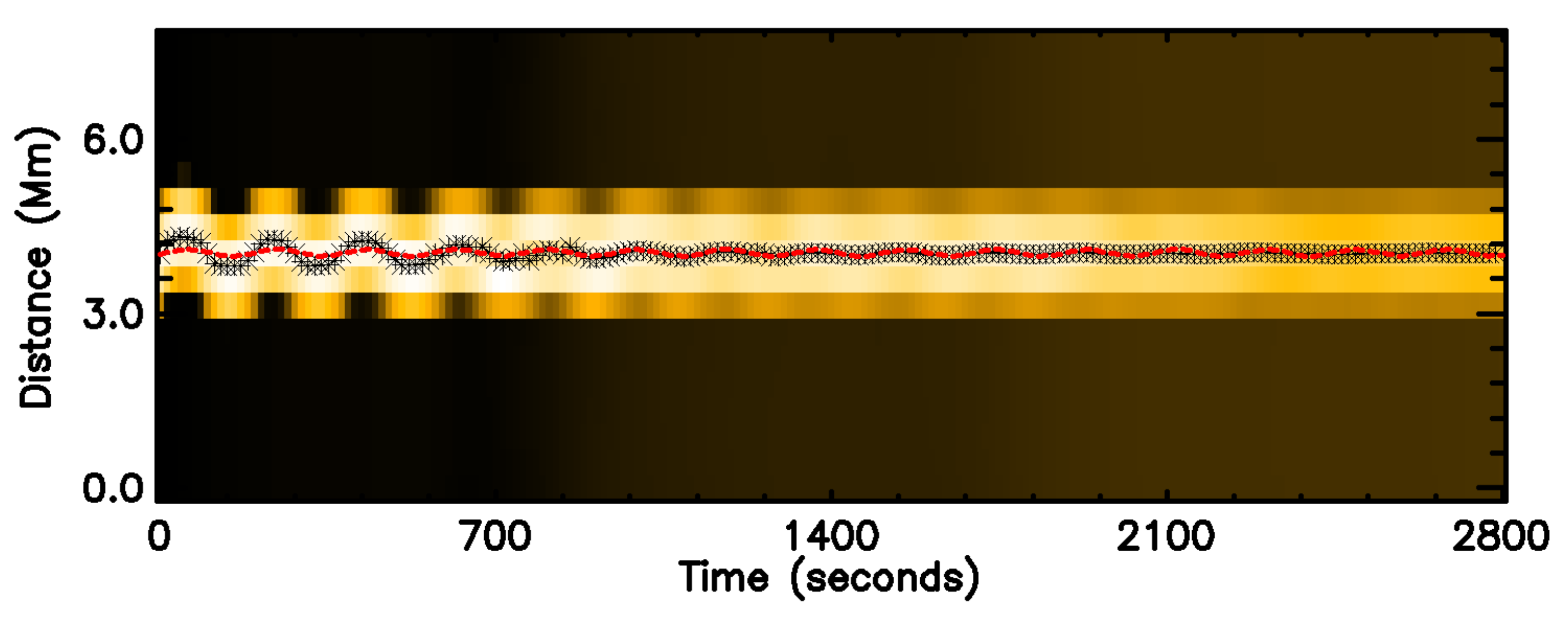}
    \includegraphics[trim={.5cm 4.cm 0.cm 0.cm},clip,scale=0.5]{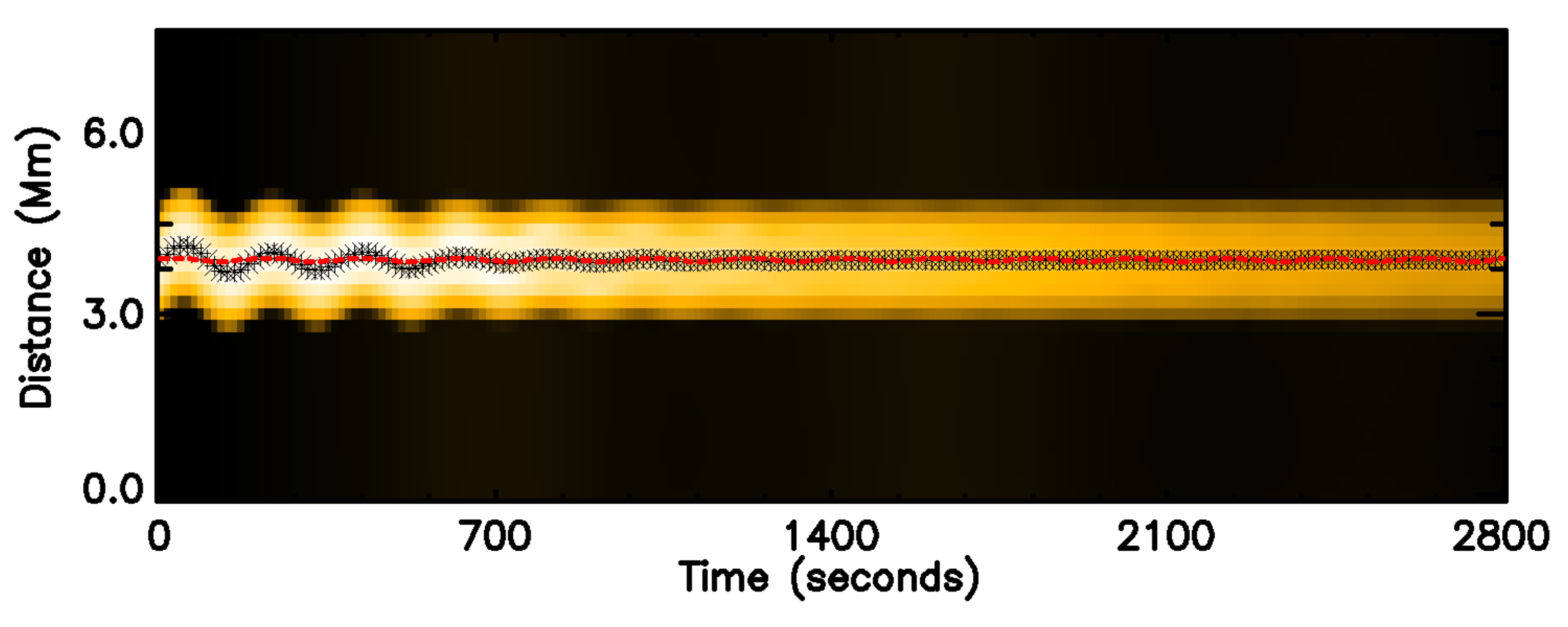}
    \includegraphics[trim={.5cm 4.cm 0.cm 0.cm},clip,scale=0.5]{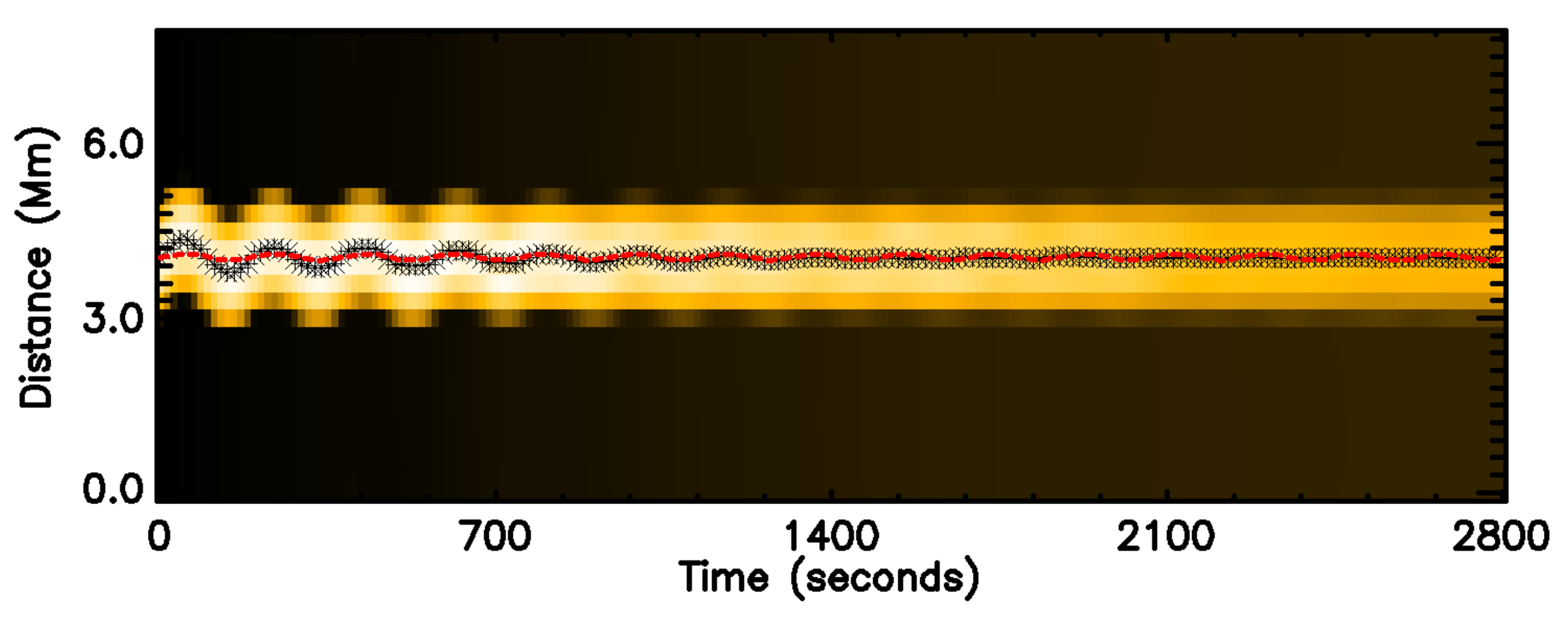}
    }
    \resizebox{\hsize}{!}{
    \includegraphics[trim={0.cm 0.cm 0.cm 0.cm},clip,scale=0.5]{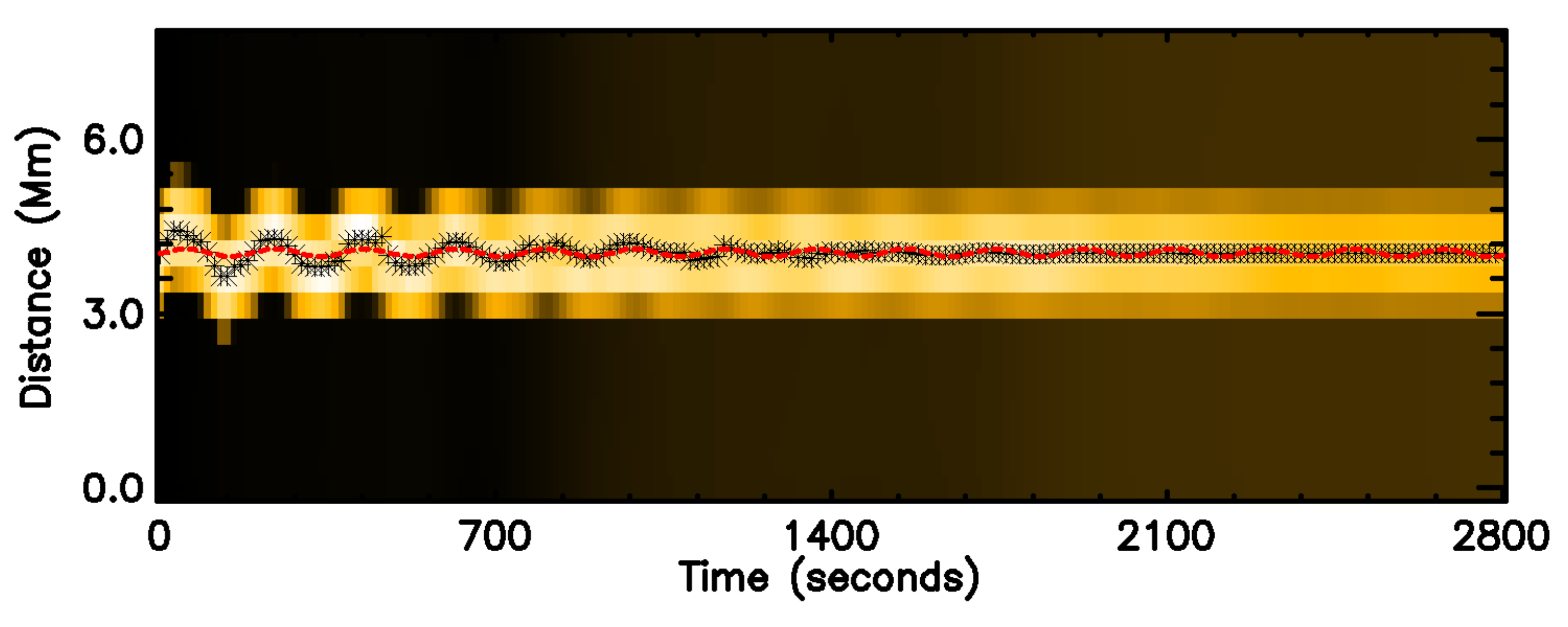}
    \includegraphics[trim={.5cm 0.cm 0.cm 0.cm},clip,scale=0.5]{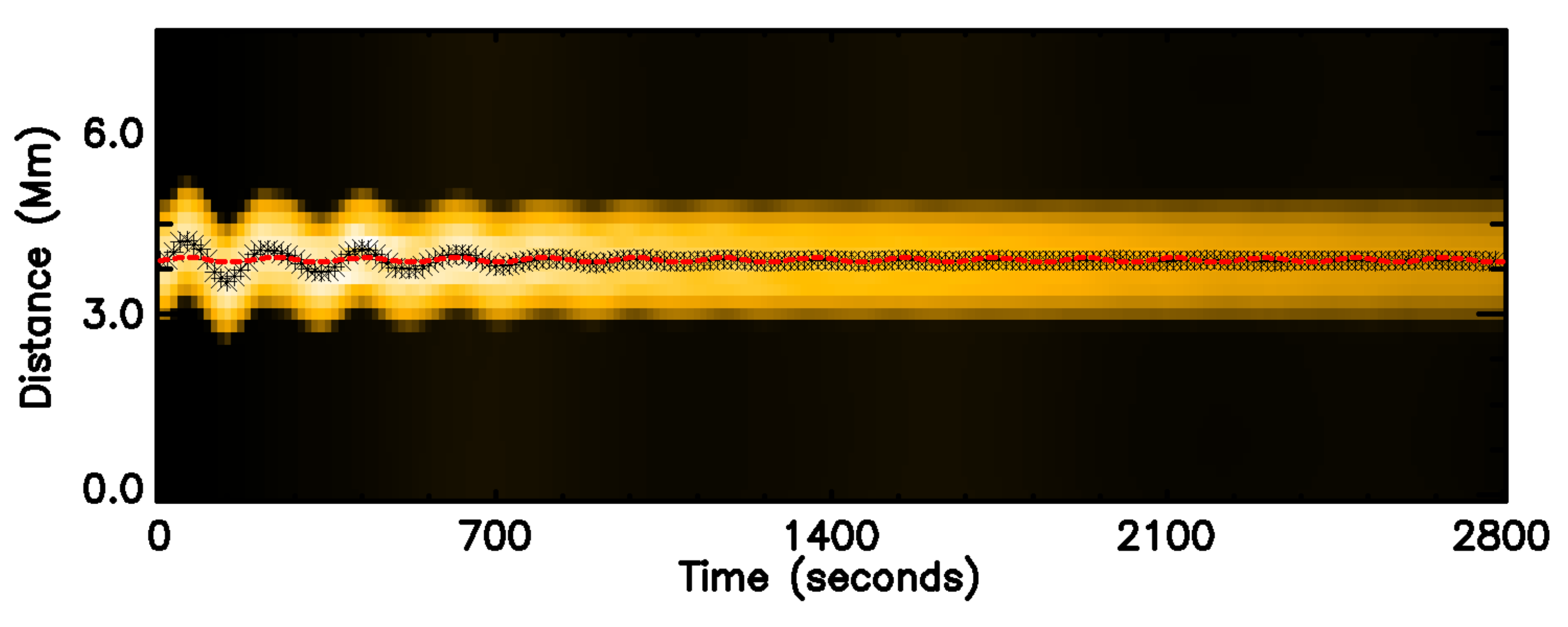}
    \includegraphics[trim={.5cm 0.cm 0.cm 0.cm},clip,scale=0.5]{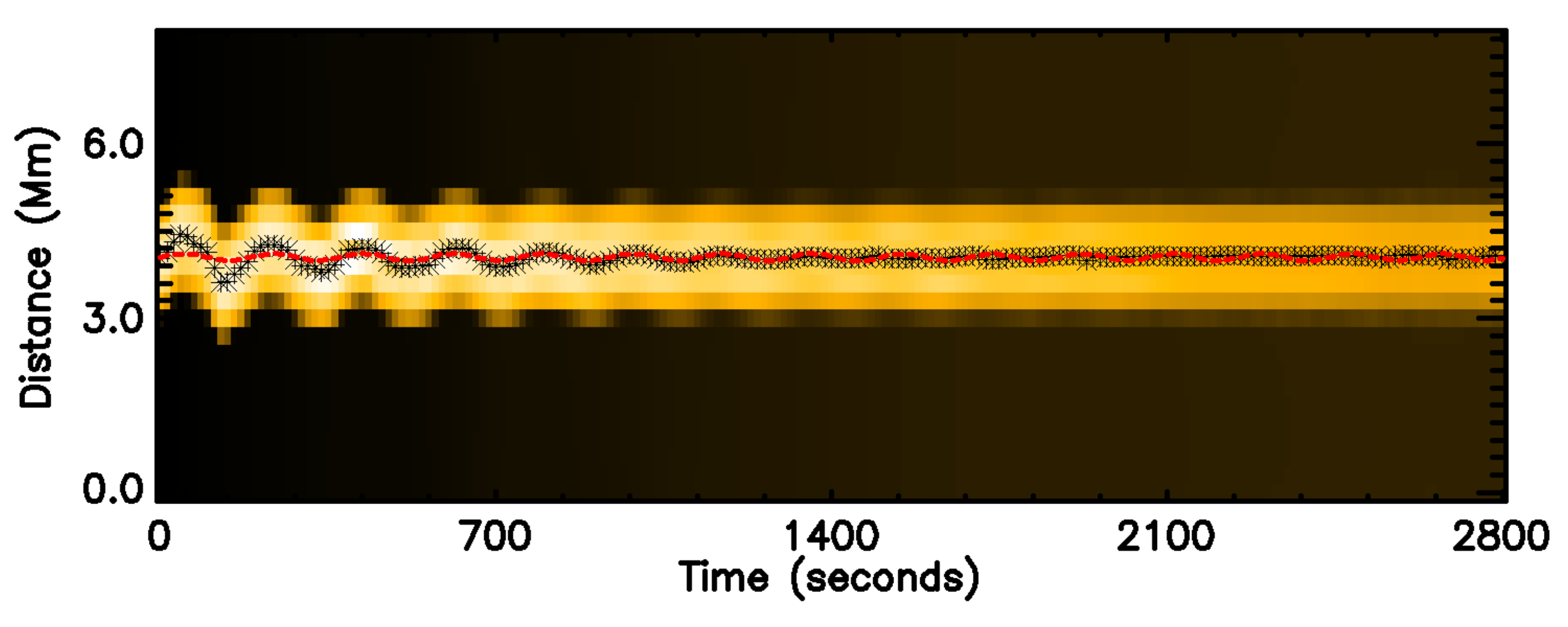}
    }
    \caption{Time-distance maps of the synthetic emission in the 171 \r{A} line, for the three cases of oscillating loops, at the apex. Left column: SDO/AIA resolution. Middle column: Solar Orbiter/HRI$_\mathrm{EUV}$ resolution.  Right column: MUSE/SG resolution. Top panels: Loop with the broadband driver. Middle panels: Loop with the sinusoidal, symmetric pulse.  Bottom panels: Asymmetric, off-centre Gaussian pulse. Overplotted in each panel are the oscillating signal of the loop centre (black asterisks) and a fitted sinusoidal function (solid red line) with the four prominent frequencies for each signal, as detected by Auto-NUWT.}
    \label{fig:tdmaps}
\end{figure*}

\begin{figure*}
    \centering
    \resizebox{\hsize}{!}{
    \includegraphics[trim={0.0cm 1.1cm 3.8cm 0.5cm},clip,scale=0.5]{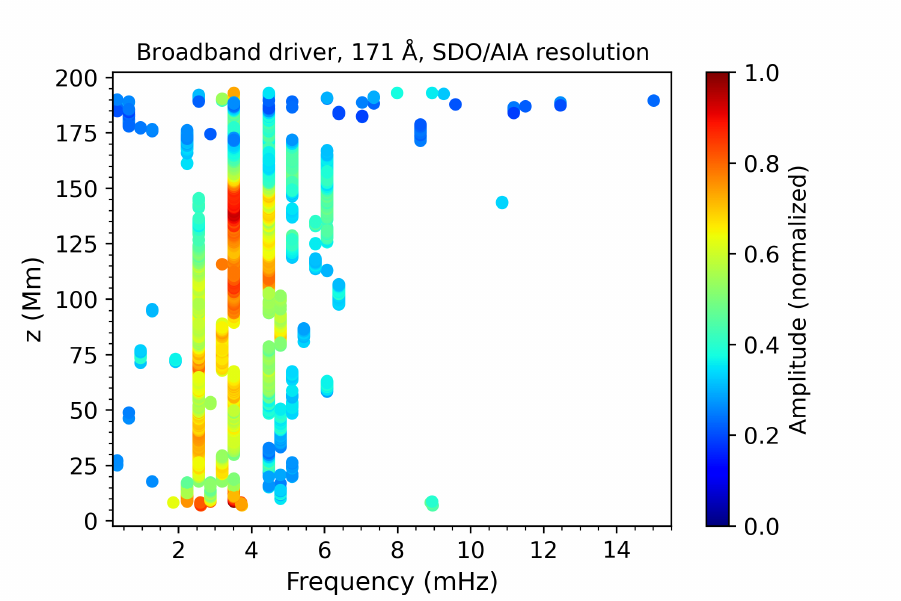}
    \includegraphics[trim={1.7cm 1.1cm 3.8cm 0.5cm},clip,scale=0.5]{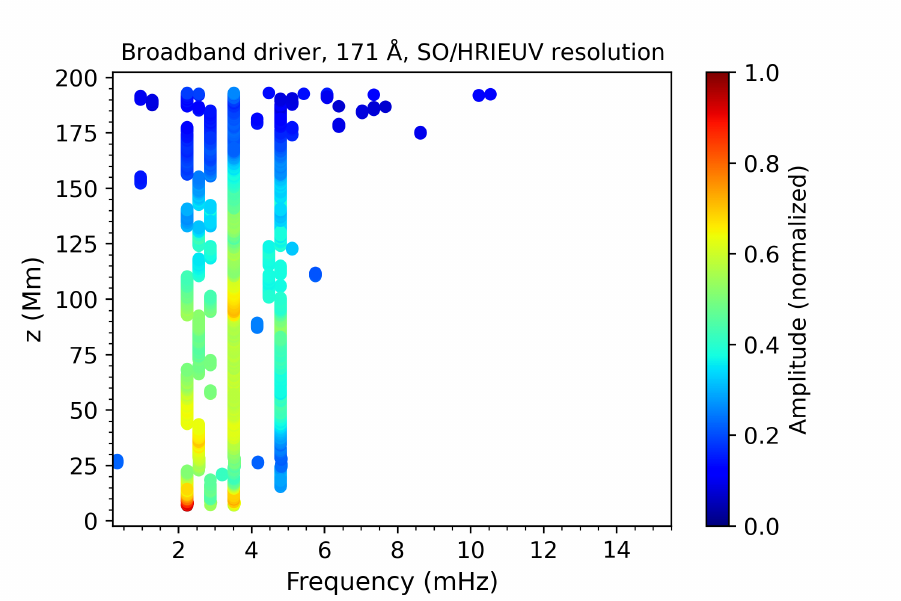}
    \includegraphics[trim={1.7cm 1.1cm 1.0cm 0.5cm},clip,scale=0.5]{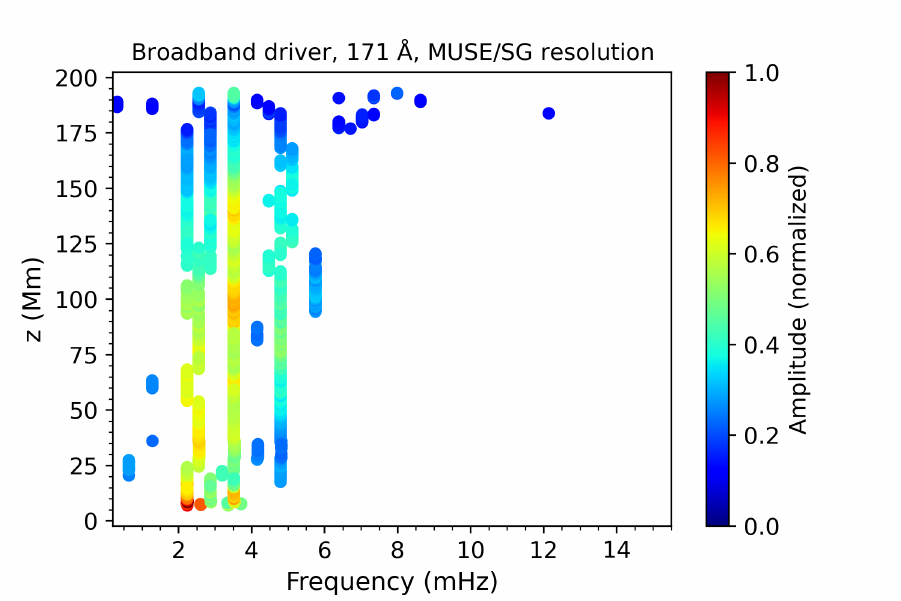}}
    \resizebox{\hsize}{!}{
    \includegraphics[trim={0.0cm 1.1cm 3.8cm 0.5cm},clip,scale=0.5]{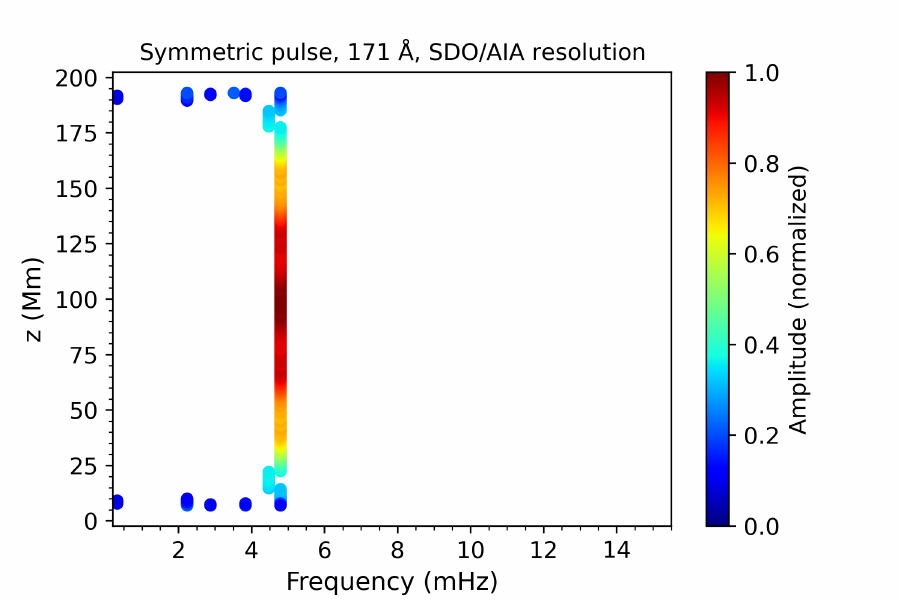}
    \includegraphics[trim={1.7cm 1.1cm 3.8cm 0.5cm},clip,scale=0.5]{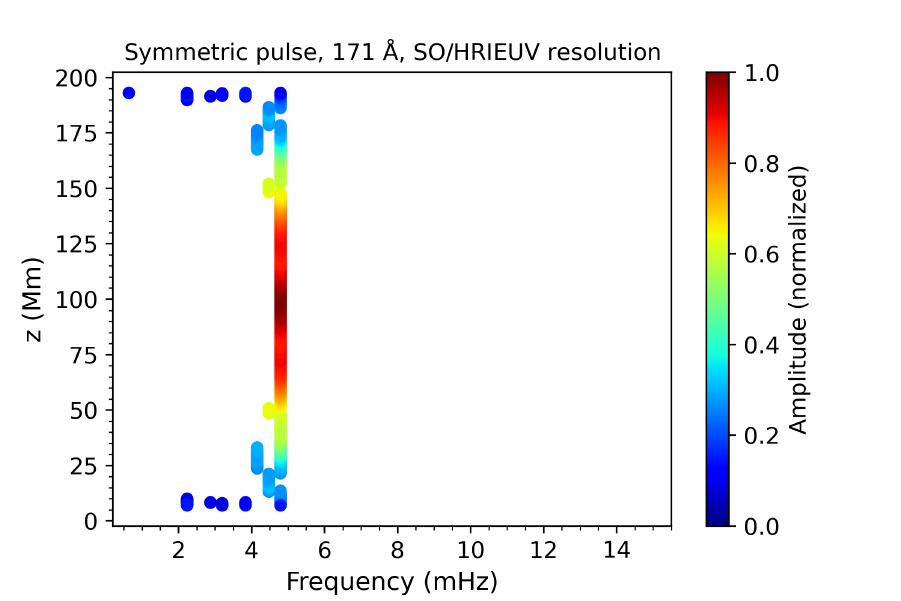}
    \includegraphics[trim={1.7cm 1.1cm 1.0cm 0.5cm},clip,scale=0.5]{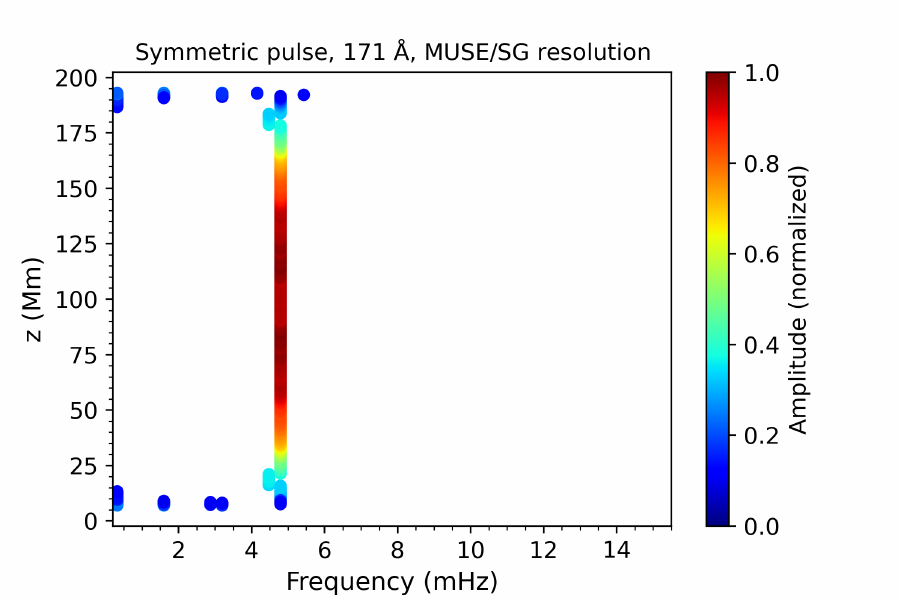}}
    \resizebox{\hsize}{!}{
    \includegraphics[trim={0.0cm 0.cm 3.8cm 0.5cm},clip,scale=0.5]{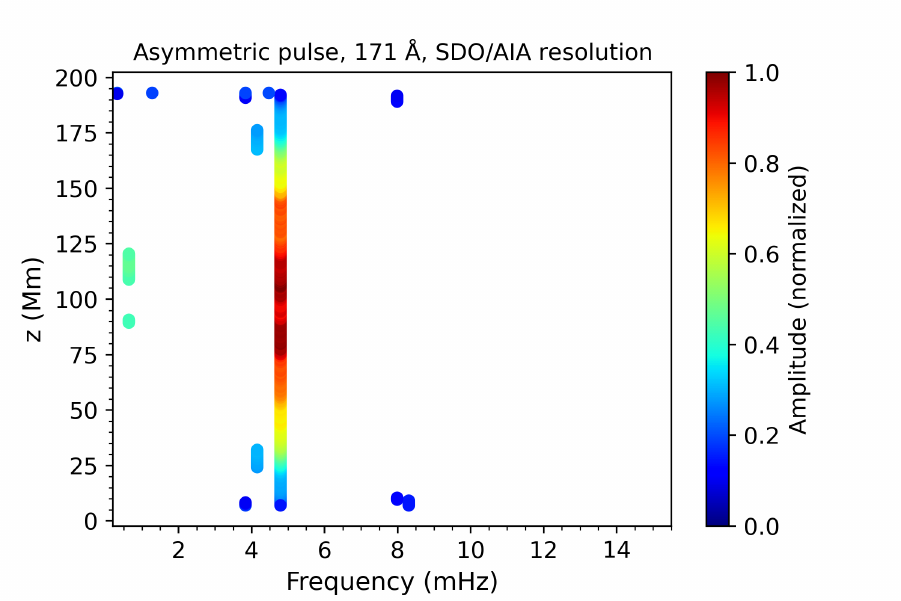}
    \includegraphics[trim={1.7cm 0.cm 3.8cm 0.5cm},clip,scale=0.5]{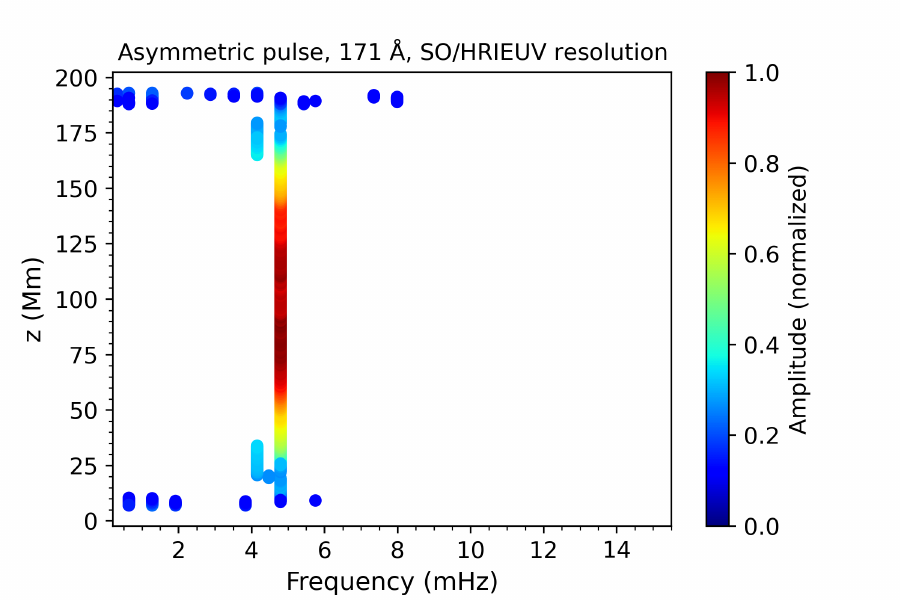}
    \includegraphics[trim={1.7cm 0.cm 1.0cm 0.5cm},clip,scale=0.5]{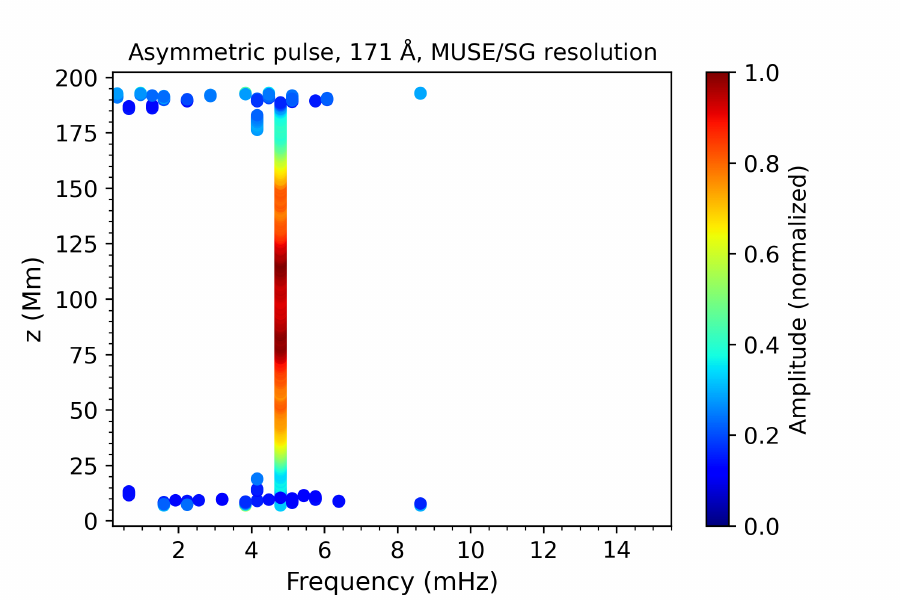}}
    \caption{Distribution of oscillation frequencies, detected with Auto-NUWT at each height, $z,$ from the synthesised 171 \r{A} line emissions. The colour bars correspond to the normalised amplitude of the oscillation for each detected frequency. As in Fig. \ref{fig:tdmaps}, the left, middle, and right columns correspond to SDO/AIA, Solar Orbiter/HRI$_\mathrm{EUV}$, and MUSE/SG resolution, while the top, middle, and bottom panels correspond to the simulations with the broadband driver, symmetric pulse, and asymmetric pulse, respectively.}
    \label{fig:NUWT}
\end{figure*}

\begin{figure*}
    \centering
    \resizebox{\hsize}{!}{
    \includegraphics[trim={0.cm 0.cm 0cm 0.cm},clip,scale=0.52]{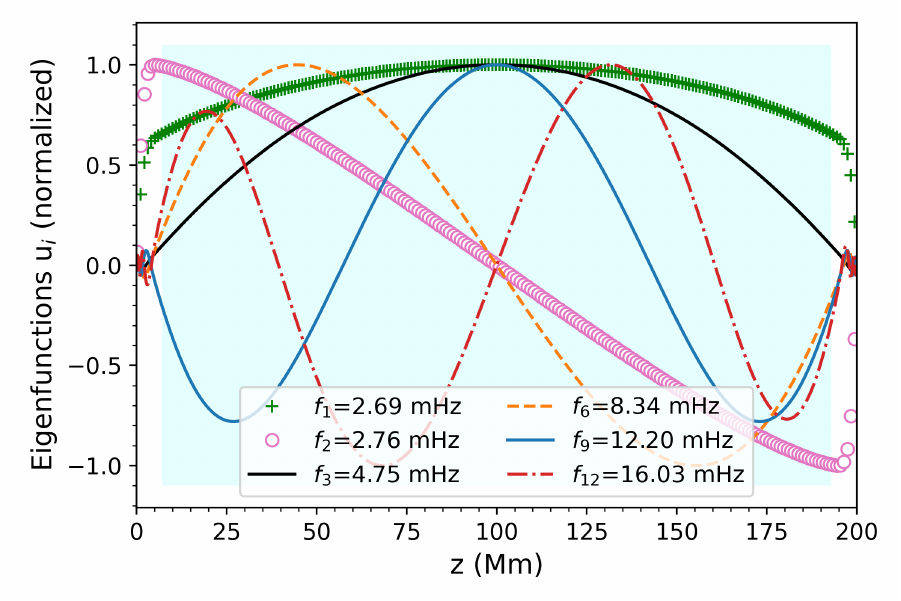}
    \includegraphics[trim={0.cm 0.cm 0cm 0.cm},clip,scale=0.52]{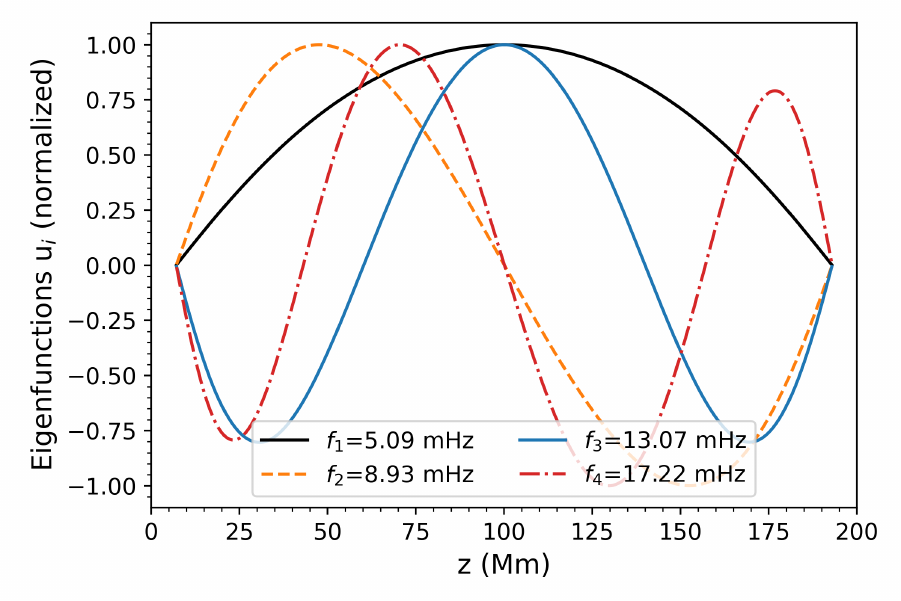}
    }
    \caption{Normalised eigenfunctions of the 1D oscillating string system and its corresponding frequencies. Left: Solution for the entire system. Right: Solution when only the coronal part of the system is used. The cyan shaded region highlights the coronal part in the left panel.}
    \label{fig:modes1D}
\end{figure*}

\begin{figure*}
    \centering
    \resizebox{\hsize}{!}{
    \includegraphics[trim={0.cm 1.1cm 0cm 0.cm},clip,scale=0.52]{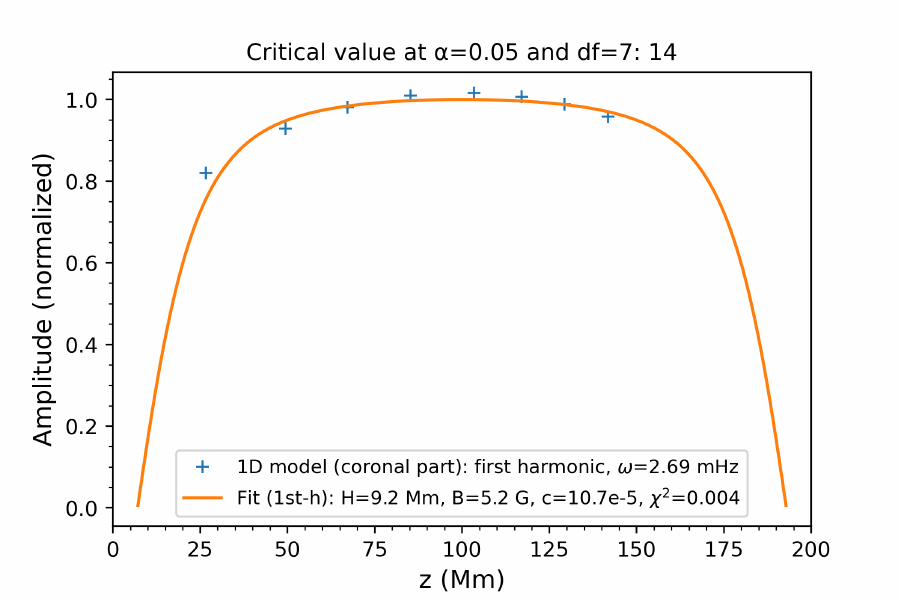}
    \includegraphics[trim={0.cm 1.1cm 0cm 0.cm},clip,scale=0.52]{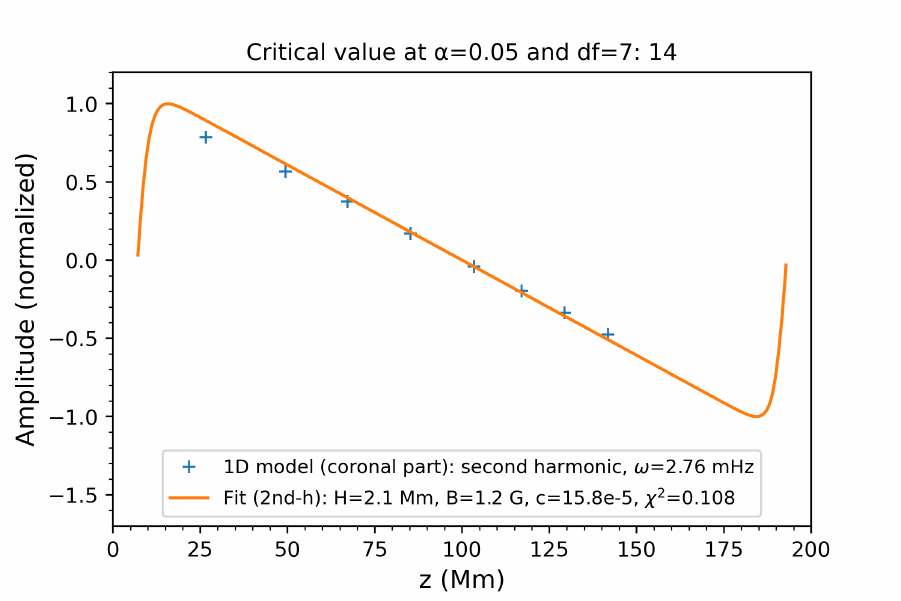}
    }
    \resizebox{\hsize}{!}{
    \includegraphics[trim={0.cm 1.1cm 0cm 1.1cm},clip,scale=0.52]{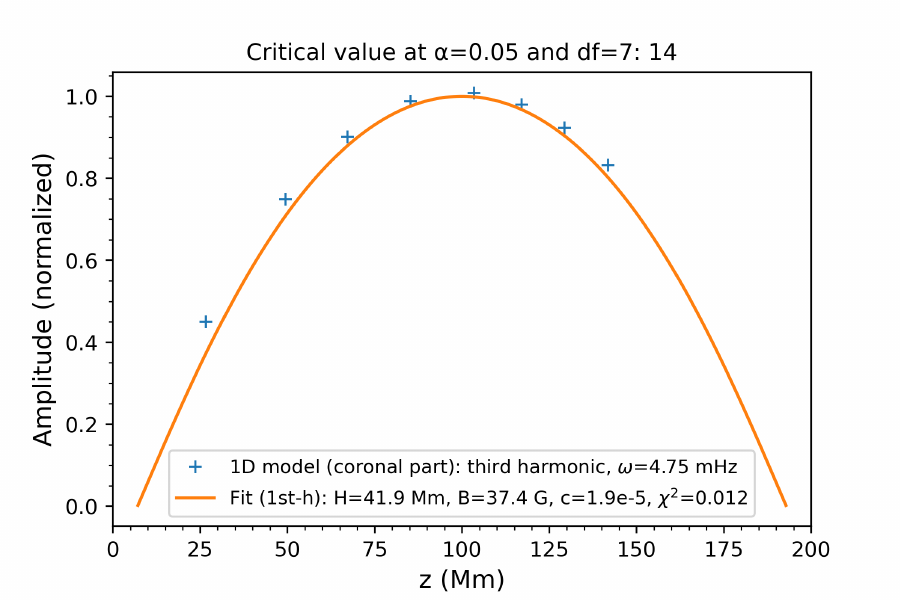}   
    \includegraphics[trim={0.cm 1.1cm 0cm 1.1cm},clip,scale=0.52]{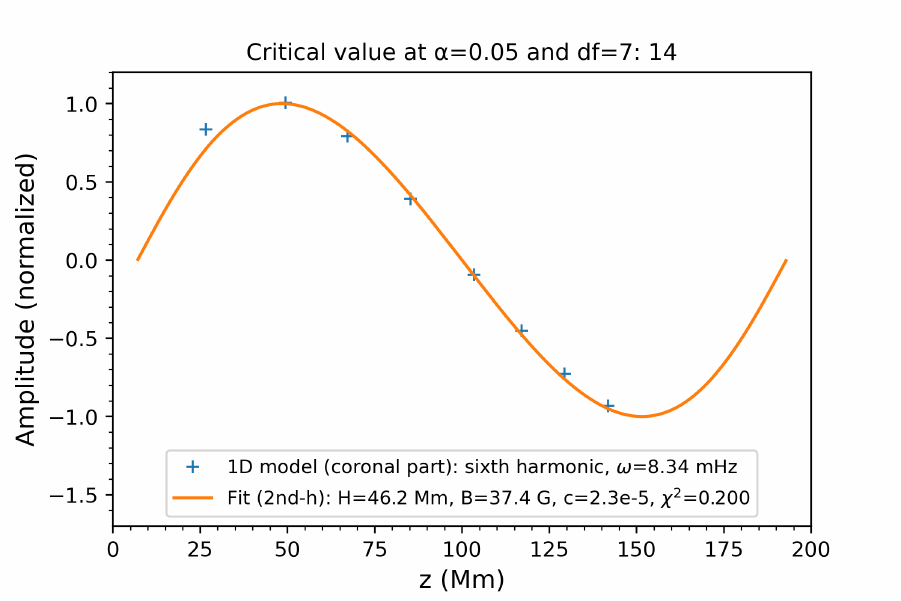}
    }
    \resizebox{\hsize}{!}{
    \includegraphics[trim={0.cm 0.cm 0cm 1.1cm},clip,scale=0.52]{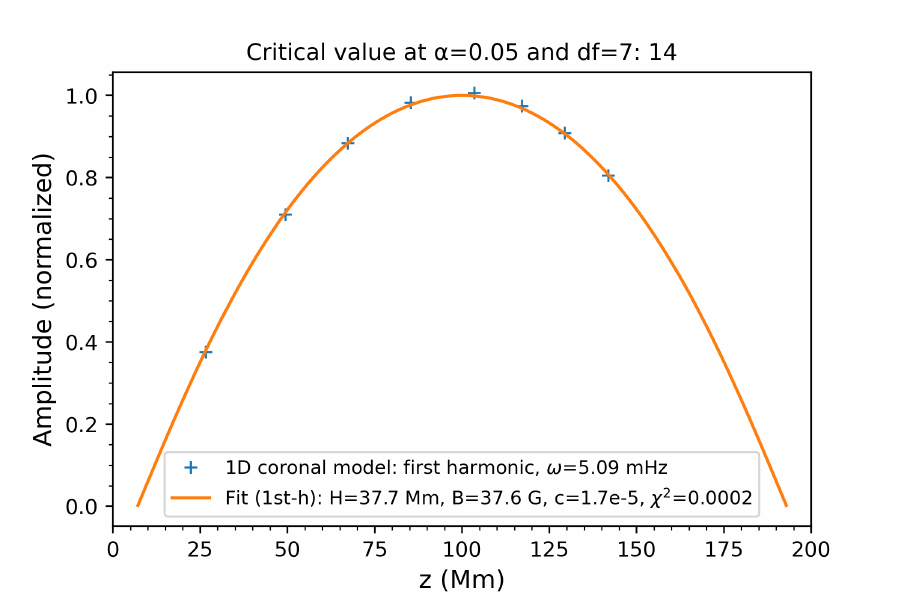}
    \includegraphics[trim={0.cm 0.cm 0cm 1.1cm},clip,scale=0.52]{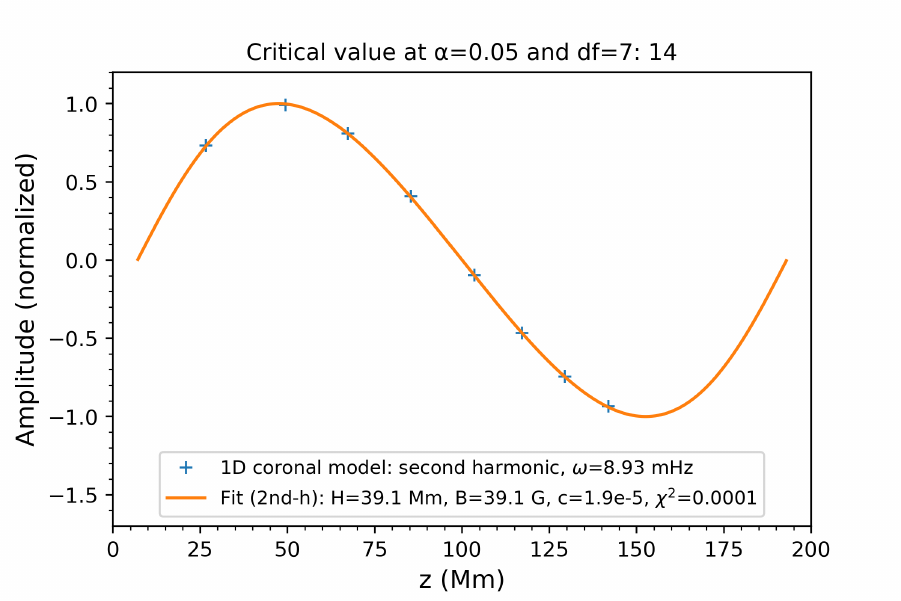}
    }
    \caption{Successful tests of $\chi^2$ goodness of fit between the eigenfunctions of the 1D oscillating string system (blue crosses, observations) and the eigenfunctions of the 1D system using a coronal model for the density and magnetic field distribution (solid orange line, prediction). Our models have a $\chi^2$ critical value of $\chi^2_\mathrm{crit}=14$ at $\alpha=0.05$ ($95\%$ significance level) with seven degrees of freedom. Bottom panels: Purely coronal system. Top and middle: Full-length 1D system. Shown are observations fit with the first harmonic of our predicted 1D model (left column) and the second harmonic (right). The first, second, third, and sixth harmonic are used as observations. The predicted scale height, magnetic field, scaling parameter, and $\chi^2$ test result are included in each panel.}
    \label{fig:chi2}
\end{figure*}

\begin{figure*}
    \centering
    \resizebox{\hsize}{!}{
    \includegraphics[trim={0.cm 0.cm 1cm 0.cm},clip,scale=0.57]{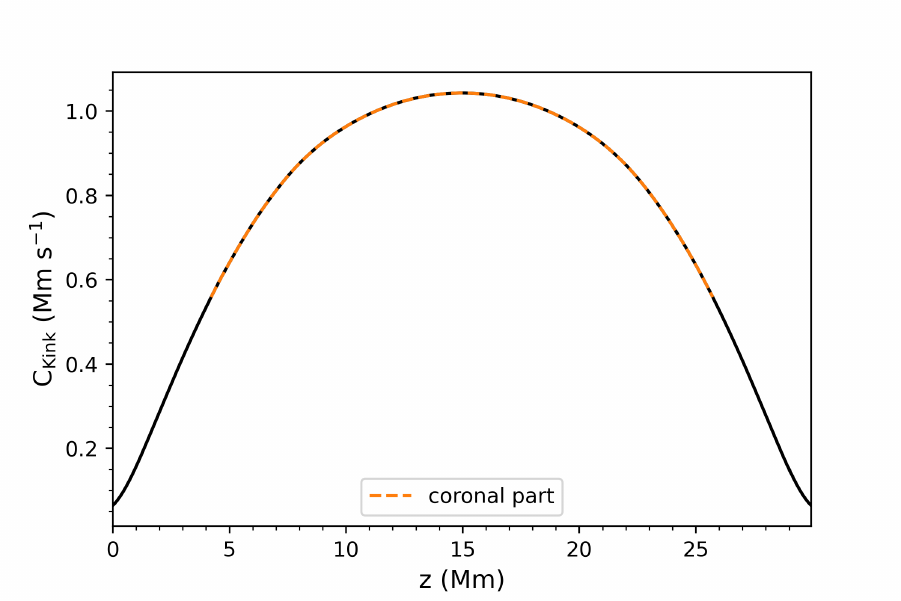}
    \includegraphics[trim={0.cm 0.cm 0cm 0.cm},clip,scale=0.52]{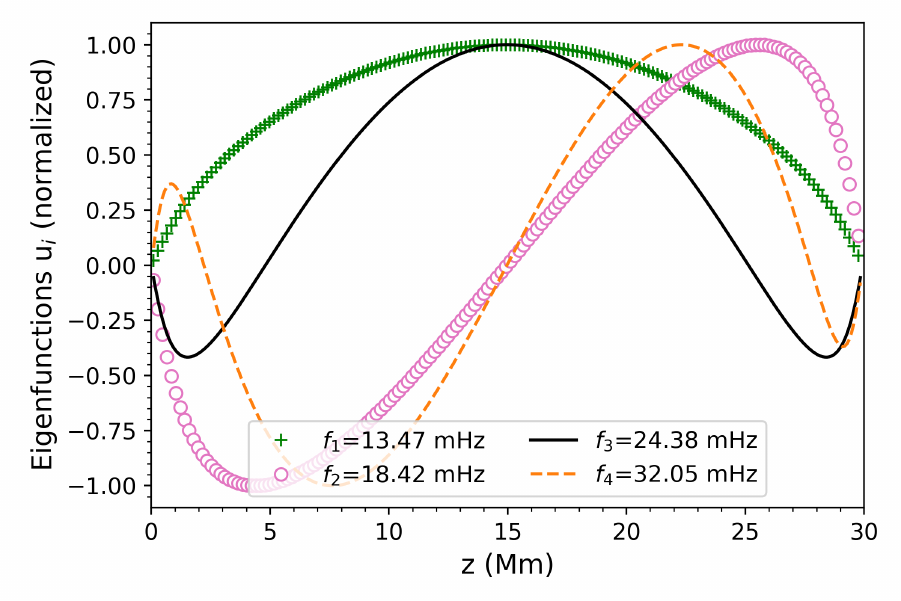}
    }
    \caption{Profiles of the kink speed along the $z$-axis (left) and the corresponding normalised eigenfunctions of the 1D system of Eq. \ref{eq:1D} (right) for the short loop model used in \citet{Gao2023ApJ...955...73G}.
    The coronal part of the kink speed profile is highlighted in the left panel ($z\in \left[ 4.2, 25.8 \right]$\,Mm, dashed orange line) and the eigenfunctions for the entire loop length in the right.}
    \label{fig:modesYuhang1D}
\end{figure*}

\begin{figure*}
    \centering
    \resizebox{\hsize}{!}{
    \includegraphics[trim={0.cm 0.cm 0cm 0.cm},clip,scale=0.52]{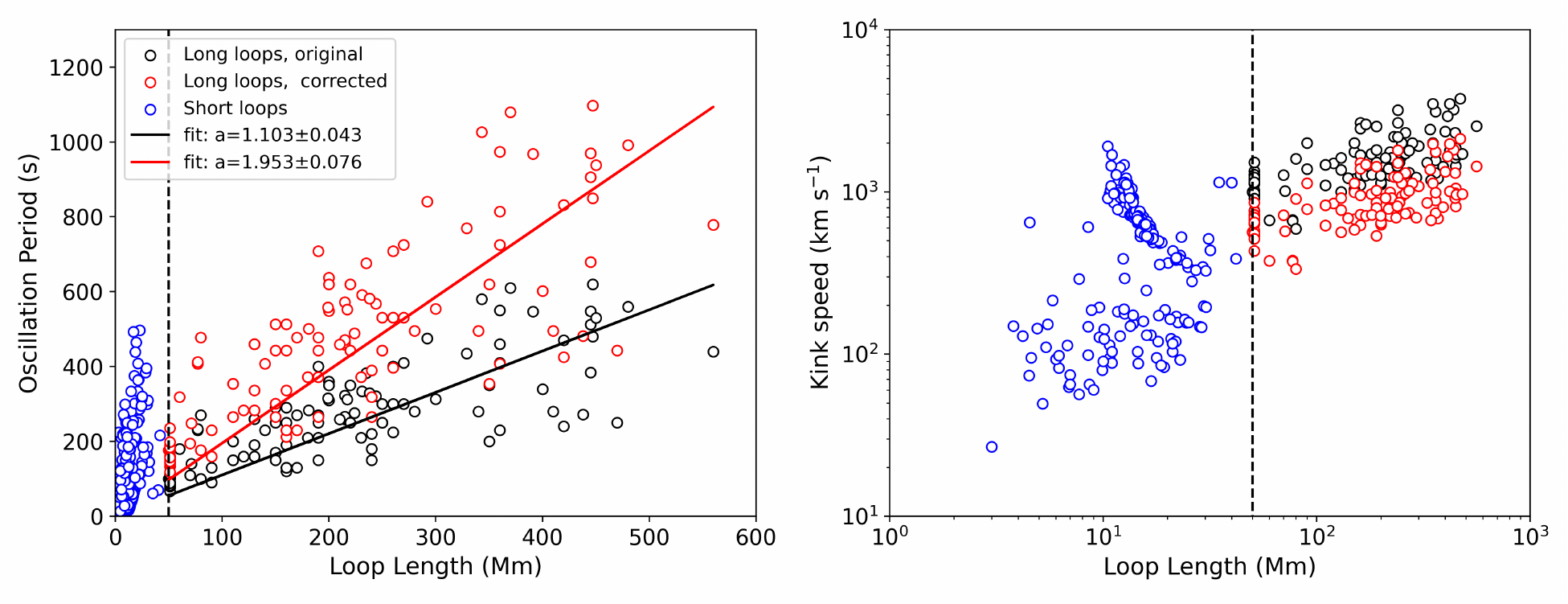}
    }
    \caption{Scatter plots for various parameters for observations of loops undergoing decayless oscillations, as reported in past works \citep{wang2012,nistico2013,anfinogentov2013,anfinogentov2015,anfinogentov2019ApJ,ZhongSihui2022MNRAS.513.1834Z,ZhongSihui2022MNRAS.516.5989Z,GaoYuhang2022ApJ...930...55G,GaoYuhang2024A&A...681L...4G,LiandLong2023ApJ...944....8L,Petrova2023ApJ...946...36P,ShrivastavArpitKumar2024A&A...685A..36S}. Left: Relation between the period and the loop length, as well as the best linear fit for the long loops. Right: Relation between the kink speed and the loop length. The dashed lines at $50$\,Mm divide the short- and long-loop data. The blue circles correspond to the short loops, the black circles to the long loops (original data), and the red circles to the long loops, after adjusting the data for measuring the third harmonic.}
    \label{fig:decayless}
\end{figure*}

This numerical study is based on the analysis of our setup that consists of a stratified 3D straight magnetic flux tube, modelling a coronal loop with chromospheric footpoints. The numerical setup is similar to the ones in \citet{Karampelas2024A&A...681L...6K} and \citet{Karampelas2024A&A...688A..80K}. Following our past work,  we initialised a 2D model in hydrostatic equilibrium along the vertical $z$ direction, using sinusoidal gravity ($g_z(z) = 274\, \cos(\pi\,z/200)$ in m s$^{-2}$), to emulate the gravity variation along a semi-circular loop. We ran a 2.5D resistive MHD simulation to let our model reach a quasi-equilibrium state and we then interpolated into a 3D domain. 

The initial profiles for the magnetic field ($\mathbf{B}$), density ($\rho$), and temperature ($T$) are
\begin{align}
    &\lbrace B_r, B_\theta, B_z \rbrace = \lbrace 0, 0, 30 \rbrace \, \text{G}\\
    &T(r,z) = T_\mathrm{Ch} + (T_\mathrm{C}(r) - T_\mathrm{Ch})( 1 - \left[ (L-z)/(L-\Delta_\mathrm{Ch}) \right]^2 )^{0.3},\\
    &T_\mathrm{C}(r) = T_\mathrm{C,e} + (T_\mathrm{C,i} - T_\mathrm{C,e})\, \zeta(r),\\
    &\rho_\mathrm{Ch} = \rho_\mathrm{Ch,e} + (\rho_\mathrm{Ch,i}-\rho_\mathrm{Ch,e})\, \zeta(r),\\
    &\zeta(r) =0.5\left[ 1 - \tanh\left(\left( \left[r/R\right]-1 \right) 20 \right) \right], \label{eq:zeta}
\end{align}
where the subscripts $\mathrm{C}$ and $\mathrm{Ch}$ correspond to the coronal ($z=100$\,Mm) and chromospheric ($z=0,200$) values. The temperature in the transition region and the corona is calculated for height $z \in \left[5, 195\right]$\,Mm. The loop length is $L=200$\,Mm, while $\Delta_\mathrm{Ch}=5$\,Mm is the width of our chromosphere. For $z<5$ Mm and $z>195$ Mm, we considered a uniform chromospheric temperature of $T_\mathrm{Ch}=0.02$\,MK. The temperature at the apex ($T_\mathrm{C}(r)$) is $T_\mathrm{C,i} = 1$\,MK and $T_\mathrm{C,e} = 1.5$\,MK inside and outside the flux tube, respectively. The density is $\rho_\mathrm{Ch,i} = 3.51 \times 10^{-8}$\,kg m$^{-3}$ inside the footpoints and $\rho_\mathrm{Ch,e} = 1.17 \times 10^{-8}$\,kg m$^{-3}$ outside the footpoints. The function $\zeta(r)$ calculates the loop cross-section, with $R=1$\,Mm being the loop minor radius.

For the 2.5D domain ($r\in[0,8]$\,Mm and $z\in[0,200]$\,Mm), we took $200\times 2048$ uniformly spaced grid points in cylindrical coordinates. We considered axisymmetry at $r=0$ and open boundary conditions at $r=8$\,Mm. At the $z$-axis boundaries, we used zero-gradient conditions for the magnetic field components, anti-symmetric conditions for the velocity components, symmetric conditions for the density, and a constant temperature. We then let the setup reach a semi-equilibrium state for a total time of $t=3890$\,s, reducing the velocity values per iteration as $v_{\lbrace r,\theta,z \rbrace} = v_{\lbrace r,\theta,z \rbrace}/1.0001$ for $t\in [778, 3112]$\,s, similarly to \citet{pelouze2023A&A...672A.105P} and \citet{mingzhe2023ApJ...949L...1G}. Figure \ref{fig:inicon} shows the post relaxation 2D temperature profile as well as 1D profiles for the density and magnetic field inside ($i,\,r=0$) and outside ($e,\,r=8$ Mm) the flux tube. After the relaxation, the magnetic field has been restructured through advection and dissipation and is no longer purely vertical, but also has a radial component ($B_r$) that has its a maximum value near the loop footpoints. This maximum value of the $B_r$ is an order of magnitude lower than the vertical component ($B_z$) there. The vertical component gets its minimum value near $r=0$ at the footpoint, and its maximum value at $r\sim 1.5$ Mm near the footpoint, as was shown for a similar setup in \citet{Karampelas2024A&A...688A..80K}. At larger radii, as well as towards the transition region and the corona, the radial component is almost zero, and the magnetic field is near-vertical there. Similarly to \citet{Karampelas2024A&A...688A..80K}, the loop minor radius does not change post relaxation, but is approximately $R=1$\,Mm, with the exception of the footpoints ($z=0,\,200$\,Mm) where it has increased to $R=1.2$\,Mm. Finally, in Fig. \ref{fig:inicon} we also plot the calculated kink speed ($C_\mathrm{Kink}$) along the $z$-axis, with the coronal part ($z\in[7,193]$\,Mm) highlighted. 

We interpolated the post 2.5D relaxation slice into a Cartesian grid of size $x\in [-4,4]$\,Mm, $y\in [0,4]$\,Mm and $z\in [0,200]$\,Mm, with $\delta x = \delta y = 0.040$\,Mm everywhere, $\delta z = 0.098$\,Mm for $z\leq 10$\,Mm and $z\geq 190$\,Mm and $\delta z = 0.8$\,Mm for $24\,\text{Mm}\leq z\geq 176$\,Mm. We considered stretched grids with $40$ cells each for $10\,\text{Mm}\leq z\leq 24$\,Mm and $176\,\text{Mm}\leq z\leq 190$\,Mm. 

For the 3D cases, we took open boundary conditions at $x=-4,4$\,Mm and $y=4$\,Mm and reflective boundary conditions at the $y=0$, which allowed us to simulate  only half of the loop. Along the $z$-axis, we considered symmetric boundary conditions for both density and pressure, as well as zero-gradient conditions for the magnetic field and anti-symmetric boundary conditions for the $v_x$, $v_y$ and $v_z$ velocity components, apart from the case where a velocity driver is employed for $v_x$ and $v_y$ at $z=0$, for $t\leq 202$\, s.

Two different cases are considered in the numerical part of this study. For the first case, we used a model similar to that in \citet{Karampelas2024A&A...681L...6K}, where at $z=0$ for $t>202$\,s, we applied a broadband, linearly polarised velocity driver:
\begin{align}
\lbrace v_x,v_y\rbrace = \lbrace V(t)\,\zeta(r_d,t),0\rbrace,
\end{align}
with the function $\zeta(r)$ from Eq. \ref{eq:zeta} calculating the driver cross-section and $r_d =\sqrt{(x - x_0(t) )^2 + y^2}$ being the location of the driver centred at $x_0(t)$ with radius $R=2.5$\,Mm. The velocity signal $V(t)$ follows a red noise spectrum ($\sim f^{-1.66}$, where $f$ is the frequency), constructed through the \textit{colorednoise 2.1.0} python package \citep[see also][]{afanasyev2020decayless}, but with reduced power at the lower frequencies ($<1.5$\,mHz) after using a high-pass filter, similar to that used in \citet{Karampelas2024A&A...688A..80K}. The left panels of Fig. \ref{fig:driver} show the velocity driver profile and its respective spectrum. The RMS velocity ($V_{RMS}$) of the signal ($\sim 0.94$\,km s$^{-1}$) is comparable to the RMS velocities of the horizontal motions of magnetic bright points ($\sim 1.32$\,km s$^{-1}$) derived from the Swedish 1 m Solar Telescope (SST) and Hinode observations \citep{Chitta2012ApJ...752...48C}, which could be treated as loop footpoints motions. 

For the second case, we ran two simulations in which we impulsively perturbed the flux tube with an initial condition in the $v_x$ velocity. We used a sinusoidal pulse, which is symmetric with respect to the apex and an off-centre Gaussian pulse, which was asymmetric with the respect to the apex:  
\begin{align}
 &v_x = 15\,\zeta(r)\,\sin{(\pi\,z/L)},\\
 &v_x = 45\,\zeta(r)\,\exp{((z-80)/30)^2}.
\end{align}
Both velocity pulses are given here in km s$^{-1}$, and $z$ is in Mm. The radius of the velocity pulse cross-section, defined by $\zeta(r)$ from Eq. \ref{eq:zeta}, is $R=1.2$\, Mm to ensure that the entire flux tube cross-section is perturbed. We note that a different amplitude was used so that the initial displacement of the flux tube would be of the same order.  In the right panel of Fig. \ref{fig:driver}, we see the velocity initial condition used to perturb our flux tube in the two simulations of that setup.

We solved the compressible MHD equations for a hydrogen plasma using the PLUTO code \citep{mignonePLUTO2007}. We enforced the $\nabla\cdot\mathbf B=0 $ condition through a hyperbolic divergence cleaning method. We used anisotropic thermal conduction, with the thermal conduction coefficient (in J\,s$^{-1}$\,K$^{-1}$\,m$^{-1}$) from the \textit{Spitzer} conductivity \citep{Orlando2008ApJ}, and the method for artificially broadening the transition region during the simulation, described in \citet{LinkerEtAl2001}, \citet{LionelloEtAl2009}, and \citet{MikicEtAl2013}:
\begin{align} 
&\kappa_{\parallel} = 5.6 \times 10^{-12} \left \{
    \begin{array}{ll}
    T^{5/2}, & \text{if $T\geq 2.5\times10^5$\,K}, \\
    (2.5\times 10^5)^{5/2}, & \text{if $T\leq 2.5\times10^5$\,K}, 
    \end{array}\right.\\
&\kappa_{\perp} = 3.3 \times 10^{-21}\, n_{H}^2/(\sqrt{T}B^2).
\end{align}
Our setup has a different value for the parallel thermal conduction coefficients than the ones used in \citet{Karampelas2024A&A...681L...6K} and \citet{Karampelas2024A&A...688A..80K} in order to match the values used in \citet{mingzhe2023ApJ...949L...1G} and \citet{Gao2023ApJ...955...73G}. In the 2.5D setup, explicit magnetic diffusivity ($\eta = R_m^{-1} = 10^{-4}$) was used to improve the code stability, while none was used in the 3D case. Our numerical scheme also introduces effective numerical diffusivity, with a corresponding effective magnetic Reynolds number of $R_\mathrm{m,eff} = 10^4 - 10^5$.

\section{Results} \label{sec:results}

We ran three 3D MHD simulations, in which our stratified loop performed transverse oscillations. In the first case, our loop is perturbed by a footpoint broadband and a linearly polarised driver, while in the second and third case we used a sinusoidal pulse symmetric along the loop velocity  and an off-centre Gaussian velocity pulse asymmetric along the loop, respectively. We limited our 3D study mostly to the coronal part of our system, and we treated the chromosphere as a mass reservoir of near-constant temperature. However, we focused on the importance of the large gradients in the transition region for our understanding the dynamics of the standing transverse oscillation modes.

\subsection{3D loop simulations and forward modelling}
Our reference step is to revisit and expand upon the result in our past work, as this is the basis for the upcoming analysis. In Fig. \ref{fig:cmspectra} we plot the time-distance maps of the loop displacement for the three simulations, alongside their respective spectra derived from Fourier analysis. The loop displacement was calculated by tracking the position of the loop centre of mass at each $xy$-plane through a weighted surface averaging, with $(\rho(z)-\rho_e(z))^2$ as the weight. Subtracting the external background density ($\rho_e$) and squaring reduced the effects of the background plasma in the calculation of the centre of mass. The left panels of Fig. \ref{fig:cmspectra} show the displacement for a small selection of coronal heights, while we used the entire array when calculating the spectra in the right panels. We also note that while the loop displacement is along the $x$-direction, the left panels depict the magnified signal projected on the $z$-axis for better visualisation. 

As was shown in \citet{Karampelas2024A&A...681L...6K} for a red noise driver and in \citet{Karampelas2024A&A...688A..80K} for a similar driver with less power in the lower frequencies, such footpoint drivers excite persisting transverse oscillations with a non-decaying, fluctuating amplitude. This can be seen on the top left panel of Fig. \ref{fig:cmspectra}. Examining the spectra in the top right panel, we again identify a frequency band at $\sim 5$\, mHz that spatially resembles the fundamental standing kink mode of the coronal part of the loop \citep[and was understood as such in][]{Karampelas2024A&A...681L...6K}. We also identify many higher harmonics, as well as wide frequency band at $\sim 2-3.5$\,mHz, the nature of which was not fully understood in our past work \citep[see][]{Karampelas2024A&A...681L...6K} and which was called descriptively as `half harmonic'. 

The middle panels of Fig. \ref{fig:cmspectra} show the time-distance map and spectra for the loop excited by symmetric sinusoidal velocity pulse. When considering the same sinusoidal pulse in \citet{Karampelas2024A&A...681L...6K}, wavelet spectra for the oscillation signal at the apex revealed a strong signal at $\sim 5$\,mHz and a very weak one at $\sim 2.5$\,mHz, which was one of the primary reasons why the harmonic at $5$\,mHz was understood as the fundamental mode, the other reason being its spatial distribution along the loop for the driven oscillations. In this study, we repeated this experiment, and, as shown on the middle panels of Fig. \ref{fig:cmspectra}, we calculated the spectra along the loop length and again see a very strong harmonic at $\sim 5$\,mHz that resembles the fundamental mode of the oscillating coronal part of the loop. Alongside it, we also detect a weaker signal at $\sim 2.5$\,mHz matching the previously mentioned half harmonic.

In the bottom panels of Fig. \ref{fig:cmspectra} we plot the same quantities as before but for the loop perturbed by an asymmetric, off-centre Gaussian pulse. This type of pulse was chosen to model in a more realistic way a transverse oscillation in a loop excited by a random pulse rather than aiming to excite a specific harmonic. As stated earlier, the amplitude of the velocity pulse was chosen so that the resulting oscillation has a similar amplitude to the case where the symmetric pulse was used. From the spectra, we again see that the main signal is coming from the mode around $5$\,mHz, with a much weaker signal than before seen at $2.5$\,mHz. Additionally, we also see higher harmonics at $\sim 8,\,12$ and $16$\,mHz, whose frequencies and spatial profiles make them appear as the second, third, and fourth harmonics with respect to the mode at $\sim 5$\,mHz, for a gravitationally stratified coronal loop \citep[e.g.][]{Andries2005ApJ...624L..57A,Safari2007}.

To compare our results with observations of decayless oscillations, we performed forward modelling to create synthetic observations for our 3D simulations. We used the FoMo code \citep{fomo2016} to calculate the extreme ultraviolet (EUV) emission for the Fe IX 171 \r{A} line from the optically thin coronal part of our oscillating loops. The 171 \r{A} line is ideally suited to the temperature range of the coronal part of our models, since it has its maximum formation temperature of $\log T = 5.93$. In our analysis we targeted the spatial resolution of current and upcoming instruments, specifically the SDO/AIA, the Solar Orbiter/EUI HRI$_\mathrm{EUV}$ and the Spectograph (SG) on board the Multi-slit Solar Explorer \citep[MUSE;][]{depontieu2022ApJ...926...52D...MUSE,Cheung2022ApJ...926...53C}. While SDO/AIA covers the 171 \r{A} line and MUSE/SG is designed to do so as well, Solar Orbiter/HRI$_\mathrm{EUV}$ has a channel that covers instead the 174 \r{A} line. This line, however, covers a similar temperature range as the 171 \r{A} line, so we can use the latter as a proxy in our analysis, similarly to \citet[][]{Petrova2023ApJ...946...36P}. We degraded the original spatial resolution of the synthetic images to $0''.6$ for SDO/AIA, to $0''.4$ for MUSE/SG, and to $200 \times 200$\,km$^2$ for Solar Orbiter/HRI$_\mathrm{EUV}$, assuming that Solar Orbiter is located at a distance of $0.52$\,AU from the Sun. That was the distance when the observations of high-frequency decayless oscillations discussed by \citet{Petrova2023ApJ...946...36P} took place with Solar Orbiter. We note here that the time resolution of our results ($\sim 15.52$\,s) was not changed. Therefore, caution is needed when directly comparing to Solar Orbiter/HRI$_\mathrm{EUV}$ data ($4$ s cadence) and to a lesser extent to SDO/AIA and MUSE/SG data ($\sim 12$ s). 

The time-distance maps of the 171 \r{A} line emission for the three simulations of the oscillating loops at the apex, for the resolution of the three instruments (SDO/AIA, Solar Orbiter/HRI$_\mathrm{EUV}$, and MUSE/SG), are shown in Fig. \ref{fig:tdmaps}. We note here that we synthesised observations at each height along the three loops but here only show the results at the apex. The top panels show the results for the loop perturbed by the footpoint driver, while the middle and bottom panels show the results for the impulsive oscillations excited by the symmetric Gaussian velocity pulse and the off-centre exponential, asymmetric velocity pulse. We used the Automatic Northumbria University Wave Tracking code (Auto-NUWT; \citealt{Morton2013ApJ...768...17M}; \citealt{Weberg2018ApJ...852...57W}) to track the transverse motions of our loops. The code tracks the loop position by measuring the intensity gradients. The position of each loop along the $x$-axis is overplotted on the panels of Fig. \ref{fig:tdmaps}. The different bin size used for resampling the data at each resolution can affect the readings of the intensity gradient, leading to slight differences in the measured oscillation signal from each targeted instrument. While the simulation with the broadband driver clearly shows a non-decaying, fluctuating amplitude, the oscillations excited by the two different pulses show no qualitative difference regarding the amplitude and the oscillation period. This implies that the two different pulses are expected to have similar power spectra density profiles, as was shown for the displacement of the centre of  mass in Fig. \ref{fig:cmspectra}.

Since our focus in this study is the spectra of the excited transverse oscillations, we wanted to extract the relevant information from the synthetic emission and compare it with the results in Fig. \ref{fig:cmspectra}. Using the location of the loop centre, we then used Auto-NUWT to identify the frequencies detected for the transverse oscillations through a Fourier analysis with a $95\%$ significance level. In each panel of Fig. \ref{fig:tdmaps} we show the fit of the following sum of sine functions:
\begin{equation} \label{eq:nuwt}
  x(t) = \sum_{i=1}^{4} c_i \sin(\frac{2\pi\,t}{P_i}+\phi) + x_0,   
\end{equation}
where the coefficient $c_i$ is the displacement amplitude for each respective period ($P_i$) and $x_0$ is just the position of the loop at $t=0$. As we see in Fig. \ref{fig:tdmaps}, this fit seems to not precisely track the loop centre (black asterisks) for the model with the broadband footpoint driver (top panels). This is because only the four frequencies with the highest amplitudes ($c_i$) are used for the fit in Eq. \ref{eq:nuwt}, while additional frequencies with less power are omitted. The models with the initial velocity perturbation (middle and bottom panels) show a better agreement between the fit and the detected loop centre, because they only exhibit one dominant frequency, as was also shown in Fig. \ref{fig:cmspectra}. Here, we also note that damping coefficients are not included in our analysis, as shown in Eq. \ref{eq:nuwt}.

The distribution and relative strength of the detected frequencies along each loop, for each instrument resolution is shown in Fig. \ref{fig:NUWT}. The panels here follow the layout of Fig. \ref{fig:tdmaps}. Focusing on first in the middle and bottom panels, we see that the synthetic images for the oscillations excited by the symmetric and asymmetric velocity pulses respectively exhibit very similar spectra, with one dominant frequency at $\sim 5$\,mHz and with very small differences between the different instrument resolutions. This is the same frequency band detected for these two cases in the spectra in Fig. \ref{fig:cmspectra}. Some additional frequencies with very low power are detected near footpoints, which can be attributed to noise near the bottom of the corona - top of the transition region in our setups. This is also at the tail end of the temperature range of the 171 \r{A} line. A few points are highlighted near the apex at about $0.5$\,mHz, for the asymmetric pulse case, as seen in the AIA resolution. However, this can also be attributed to noise, since no respective signal has been detected in the spectra in Fig. \ref{fig:cmspectra}. We should highlight here that no other harmonic or mode in general can be detected in the middle and bottom panels, apart from the one at $\sim 5$\,mHz. 

For the simulation with the footpoint driver, the detected frequencies from the synthetic observations are shown in the top panels of Fig. \ref{fig:NUWT}. Unlike the other two cases, multiple frequency bands are detected along the loop, including the one at $\sim 5$ mHz. Traces of higher frequency modes are present here, but the main signal comes from the frequency range of the previously mentioned half harmonic. We detect distinct frequency bands along $z$ at $\sim 2.5,\, 3.5$ and $4$ mHz. Increasing the resolutions reduces the number of frequencies with power above the $95\%$ significance level, again showing the effects that the resolution has on the detection algorithms. It is also interesting to note that while the frequency band at $5$ mHz has a power distribution along $z$ with the maximum at the apex and minima near the footpoints, the other frequency bands at lower frequencies exhibit additional strong power near the driven footpoint at $z=0$. This implies that they are at least partially associated with the driving frequencies of the footpoint driver. A similar phenomenon was observed in simulations of short loops in \citet{Gao2023ApJ...955...73G}, where the frequency of an inclined p-mode driver also manifested in wavelet spectra of the transverse oscillations alongside the fundamental mode and it harmonics. Also, in \citet{Karampelas2024A&A...681L...6K}, the addition of a different driver to the second footpoint did not modify the spectra, but seemed to reinforce these frequencies near the second footpoint.

\subsection{1D wave equation eigenvalue problem}\label{sec:1D}
To better understand the nature of the observed frequency bands in the spectra of our oscillating loops and conclude which are associated with the oscillating system itself and which with the broadband nature of the footpoint driver, we needed to calculate the natural frequencies of the standing transverse modes of our system. An analytical approach is not feasible for the scope of this study, due to the non-uniform nature of the kink speed of our system and its strong gradients along the loop. Instead, we employed the approach used in \citet{DymovaRuderman2005SoPh..229...79D}, and the approach used in \citet{HowsonBrue2023MNRAS.526..499H} for the case of Alfv\'{e}n waves. For the thin tube approximation, the modes and frequencies of non-axisymmetric, transverse oscillations of loops were calculated using the Sturm-Liouville problem
\begin{equation}
    \frac{\partial^2 u}{\partial z^2} + \frac{\omega^2}{v_\mathrm{ph}^2(z)}u=0,\quad u(z=0,L)=0.\label{eq:1D}
\end{equation}
The phase speed along the loop $v_\mathrm{ph}$ is taken equal to the kink speed $C_\mathrm{Kink}$ that we have calculated and show in Fig. \ref{fig:inicon}. We numerically solved Eq. \ref{eq:1D} to find the eigenfunctions $u$ that correspond to the standing kink mode harmonics and the eigenvalues $\omega^2 = (f/2\pi)^2$, which give the respective frequencies $f$ for each harmonic. We note here that the thin tube approximation used to derive Eq. 12 is only valid for the coronal part of our loop model. For the part of the loop below the corona, the scale height and thus also the vertical scale are expected to drop significantly from their coronal values, reducing the validity of the thin tube approximation. Therefore, the results of Eq. \ref{eq:1D} are taken as an approximation of the actual transverse modes of the system.

Figure \ref{fig:modes1D} shows the normalised transverse oscillation modes and their frequencies for our loop model. The left panel shows the solutions for our entire loop, or when the entire $C_\mathrm{Kink}$ profile from Fig. \ref{fig:inicon} was used, and the right panel shows the solutions only for the coronal part of the loop (i.e. when only the coronal part of the $C_\mathrm{Kink}$ profile was used, shown with the dashed orange line on the bottom right panel of Fig. \ref{fig:inicon}). Focusing on the left panel, we notice that the normalised eigenfunctions are deformed, with the nodes of the higher-order harmonics being pushed towards the footpoints of our loop. This effect has been  been reported in past studies of straight flux tubes and was deemed to be caused by the density stratification along the tube axis for models with a straight magnetic field along height \citep[e.g.][]{ErdelyiVerth2007A&A...462..743E,Verth2007A&A...475..341V}, that is, for loops with a non-expanding magnetic field. Similarly to \citet{Howson2022A&A...661A.144H}, we see that the existence of the transition region leads to an extreme manifestation of this effect. Considering a coronal loop rooted at the base of the corona, we get eigenfunctions with less extreme deformation with respect to the case of no density stratification, as we see in the right panel of Fig. \ref{fig:modes1D}. From the mode profiles for the two cases considered here, we see that the first two kink oscillation harmonics, which are the fundamental mode and the first overtone, are significantly deformed, with both of them exhibiting large amplitudes near the base of the corona ($z\sim 7$ and $193$ Mm). As expected, such a qualitative behaviour is absent from the coronal loop model, for which all harmonics exhibit nodes at the base of the corona. Additionally, modes like the third, sixth, ninth, and twelfth harmonics of the full loop have nodes located below, but near the base of the corona and show similar spatial profiles to the first, second, third, and fourth harmonics for the coronal loop. 

From the same two panels of Fig. \ref{fig:modes1D}, we also get the frequencies of each oscillations mode, shown in the respective legends. Focusing on the case of the full loop (left panel), we see that the fundamental mode and the first overtone have frequencies $f_1=2.69$ mHz and $f_2=2.76$ mHz, respectively. These frequencies fall within the range of the half harmonic mode that is detected in the spectra shown in Figs. \ref{fig:cmspectra} and  \ref{fig:NUWT}. These two modes exhibit very similar frequencies, due to the extreme density gradient in the transition region \citep[][]{andries2005,Andries2005ApJ...624L..57A}. The third harmonic has an eigenfrequency equal to $f_3=4.75$ mHz, which is shown to be very similar to that of the fundamental standing kink mode of the coronal part of our loop ($f_1=5,09$ mHz), as shown in the right panel of Fig. \ref{fig:modes1D}. It is also very interesting to note that the eigenfrequencies included higher harmonics for the full loop ($f_6=8.34$ mHz, $f_9=12.20$ mHz and $f_{12}=16.03$ mHz) are again similar to those of the second ($f_2=8.93$ mHz), third ($f_3=13.07$ mHz) and fourth harmonics ($f_4=17.22$ mHz) of the coronal loop, in the same way as their respective spatial profiles.

The results of the 1D analysis can thus explain the findings in our past work \citep{Karampelas2024A&A...681L...6K,Karampelas2024A&A...688A..80K} as well as in Figs. \ref{fig:cmspectra} and \ref{fig:NUWT}. What we have descriptively been referring to as half harmonic is in fact the combined result of the loop expected eigenmodes with the added effects of the continuous broadband driving, when the footpoint driver is employed. The thin frequency band around $2.5$ mHz detected for the oscillating loops without the footpoint driver correspond to the fundamental and first overtone of the system. The additional frequencies between $2.5$ and $3.5$ mHz detected for the loop with the footpoint driver are associated with the spectrum of the broadband driver itself. This is further reinforced by the results in simulations of short loops \citep{Gao2023ApJ...955...73G}, where the frequency of an inclined p-mode driver also manifested in wavelet spectra of the transverse oscillations alongside the fundamental mode and the second and third harmonics. Additionally, in simulations of a stratified but purely coronal loop perturbed with monochromatic footpoint drivers of various frequencies, the 3D flux tube would exhibit a resonant response to the monochromatic drivers, even for small deviations from the resonant frequency \citep{afanasev2019}. Although not focused on in that study, this could be the result of the non-linear evolution of the loop profile due to the development of the instabilities.

The frequency band at $\sim 5$ mHz, detected in Fig. \ref{fig:cmspectra} and in Fig. \ref{fig:modes1D} is the third harmonic of the full-loop system with frequency $f_3=4.75$ mHz. In past studies \citep{Karampelas2024A&A...681L...6K,Karampelas2024A&A...688A..80K} this mode was misidentified as the fundamental mode of the system, due to its spatial profile and as we see from the right panel of Fig. \ref{fig:modes1D}, this mode closely resembles the fundamental mode of the coronal part of our loop model. This mode gives a strong signal in the spectra from the synthetic observations of the driven model and is the dominant mode in the spectra for the oscillating loops without the footpoint driver. The prominence of this mode could therefore lead to it being misidentified with the fundamental mode, which in turn can have implications for coronal seismology, as was discussed in \citet{HowsonBrue2023MNRAS.526..499H}, as well as in \citet{Karampelas2024A&A...681L...6K} for a a system driven by a broadband footpoint driver. Our 1D analysis also helps us identify the nature of the higher harmonics detected in the panels of Fig. \ref{fig:cmspectra}, the frequencies and spatial profile of which seems to closely match that of the harmonics expected from the 1D eigenvalue system.

\subsection{Using the 1D eigenfunctions for seismology}
To explore how the effects of the density stratification can influence the results derived by seismological techniques, we employed various tools to study the results from our 1D analysis. In a recent paper by \citet{ChenErdelyi2023A&A...678A.205C}, spatial seismology methods were used to infer the density scale and magnetic field expansion factor. This was done by fitting observations of transverse oscillations of loops with the expected profiles of transverse standing oscillations modes, using a $\chi^2$ goodness of fit test. Here we employed a similar approach to calculate the density scale height and magnetic field for various cases. For the role of observations, we considered randomly selected data points ($\eta_i$,  measurements) from (a) the first, second, third and sixth harmonics of the full-loop system, as well as (b) from the first and second harmonic of the coronal loop system. For both cases, we used the results derived from the 1D eigenvalue problem of Eq. \ref{eq:1D}. As predictions $\psi$, we used the first and second harmonics from the eigenvalue problem of Eq. \ref{eq:1D}, calculated for a phase speed ($v_\mathrm{ph}(z)$) equal to the kink speed of a coronal loop model with an exponential density profile, in sinusoidal gravity like in our 3D simulations, with a straight, uniform magnetic field, $\mathbf{B}$:
\begin{align}
    &\mathbf{B} = \lbrace0,0,B_z\rbrace = \lbrace0,0,B\rbrace\\
    &\rho_{ie} = \rho_{0,ie} \exp\left(-L\sin\left(\pi z/L\right)/(\pi H)\right)\\ \label{eq:1Dfitmodel}
    &\epsilon = \left(\rho_{0,e}/\rho_{0,i}\right) = 0.1\\ 
    &v_\mathrm{ph}^2(z) = \frac{\rho_i V_{A,i}^2 + \rho_e V_{A,e}^2}{\rho_i+\rho_e} = \frac{2}{1+\epsilon}\frac{B_z^2}{\mu_0\rho_i}.
\end{align}
Here, the subscripts `$i$' and `$e$' indicate the values at the centre of the loop cross-section and at a radial distance of $r=8$ Mm from the loop, respectively, $V_A$ is the Alfv\'{e}n speed, $L=186$ Mm is the length of the fitted coronal loop model and $\rho_{0,i} = 20.63\times 10^{-12}$ kg m$^{-3}$ is the density value inside the loop at the bottom of the corona, for the considered model. 

To fit the model, we employ the same technique as described in \citet{ChenErdelyi2023A&A...678A.205C}. We first coded an iterative stochastic global search optimisation algorithm, known as a simulated annealing algorithm. Starting from an initial guess for the density scale height ($H=40$ Mm) and the uniform magnetic field ($B_z=30$ G), we performed at each iteration a $\chi^2$ goodness of fit test:
\begin{equation}
    \chi^2 (H,c) = \sum_{i=1}^N \frac{\left( \psi(H,z_i) - c \, \eta_i \right)^2}{\psi(H,z_i)}.
\end{equation}
The coefficient $c$ is used to normalise the observations $\eta_i$ and generally has dimensions the inverse of the dimensions of the observations. As an initial guess for this coefficient we take $c = \eta_{i,max}^{-1}$. The total number of measurements are equal to $N$. At each iteration, we solve Eq. \ref{eq:1D} for randomly adjusted values of $H$ and $c$, and we use the new prediction $\psi$ to minimise the $\chi^2$ statistic. We then compare the final value of the $\chi^2$ against the critical value for the $\chi^2$ test ($\chi^2_\mathrm{crit}$), at $95\%$ significance level (or $\alpha=0.05$) with $N-1$ degrees of freedom. If $\chi^2$ is above the critical value, then we reject the null hypothesis that the observations are part of predicted values. If the critical value is above the $\chi^2$ statistic, then we fail to reject the hypothesis, and thus we consider the model to be a good fit for the observations. Once we reach a good fit, we then vary the value of the uniform magnetic field $B$, for the predicted eigenfrequency of the model to match the frequency of the observations.

The results for our set of successful $\chi^2$ fitting tests are shown in the panels of Fig. \ref{fig:chi2}. For our tests, we have a critical value $\chi^2_\mathrm{crit}=14$ for $7$ degrees of freedom ($N=8$ data points) and a $95\%$ significance level. We took data points from (a) the first, second, third and sixth harmonics of the full-loop system and (b) from the first and second harmonics of the coronal loop system. We then fit those data points with the first and second harmonics calculated for the coronal loop model with the exponential density profile, described in Eq. \ref{eq:1Dfitmodel}. 

Starting with the modes for the full loop, when fitting the observations of the third (sixth) harmonic of the full loop with the fundamental (first overtone) of the fitted 1D model gives us $H=41.9$ Mm ($H=46.2$ Mm) and $B=37.4$ G. The magnetic field in particular is very close to the mean value in our 3D models ($B_z=30$ G), showing the strength of this seismological technique. On the other hand, considering the fundamental and first overtone of the full-loop system as observations leads to vastly different results for the scale height ($H=9.2$ and $2.1$ Mm) and the magnetic field ($B=5.2$ and $1.2$ G). This big difference in the results is due to the fact that the first two harmonics of the full-loop system are the most influenced by the sharp gradients in the transition region, rendering the usual approach in seismology of fitting simple coronal model profiles problematic.

To test the validity of our $\chi^2$ goodness of fit tests, we compared our results with more traditional seismological techniques, using the formulas for the ratio of periods derived in \citet{Andries2005ApJ...624L..57A} and \citet{Safari2007}:
\begin{align}
    &P_1 = P_\mathrm{Kink}(1+L/(3\pi^2H))^{-1}, \label{eq:P1}  \\ 
    &P_2 = P_\mathrm{Kink}(2+2L/(15\pi^2H))^{-1}, \label{eq:P2} \\
    &P_3 = P_\mathrm{Kink}(3+3L/(35\pi^2H))^{-1}, \label{eq:P3}
\end{align}
where $P_\mathrm{Kink}=2L/C_\mathrm{Kink}$ is the period of the fundamental standing kink mode for a loop of the same length, mean magnetic field, and no density stratification along its axis, with density equal to its value at the footpoints \citep{edwin1983wave}. Using the modes for the full loop, the periods of the fundamental ($P_1$) and the second harmonic ($P_2$) give $P_1/P_2 = 1.03$ (for an observed loop length in coronal lines equal to $L=186$ Mm) and a corresponding scale height of $H=4.03$ Mm. The latter is the same order of magnitude to the values calculated from the $\chi^2$ tests. If we instead misidentify the third and sixth harmonic as the first and second, then we get $P_\mathrm{1,false}/P_\mathrm{2,false} = 1.76$ and $H=34.9$ Mm, which again is similar to the results of the $\chi^2$ fit tests.

Applying the same $\chi^2$ goodness of fit tests when using the first (second) harmonics of the coronal part of the loop as measurements, we then get $H=37.7$ Mm ($H=39.1$) Mm for the scale height and $B=37.6$ G ($B=39.1$ G) for the magnetic field, respectively. The results can be seen in the bottom two panels of Fig. \ref{fig:chi2}. These values for the derived scale height and the magnetic field are very close to those we get for the third and sixth harmonics of the full-loop system. Additionally, when filling in the values from the modes of the coronal part of the loop in Eqs. \ref{eq:P1}-\ref{eq:P3}, and considering the same loop length as in the case of the full loop, we get $P_1/P_2 = 1.75$ and $H=34.6$ Mm. These results further hint to the possibility of observations misidentifying the order of the observed standing kink modes, if we assume that the oscillations occurred along the entire loop length and not just the coronal part. For reference, when using the modes for the coronal part of the loop and considering the same loop length we get $P_1/P_2 = 1.75$ and $H=34.6$ Mm.

\subsection{Seismological implications for transverse oscillations}
In the context of coronal seismology, it is also useful to revisit the results from older observational studies and try to understand them in light of our current results. In \citet{duckenfield2018ApJ}, both the first and second harmonics were detected in an observation of a decayless transverse oscillation of a coronal loop with a period ratio equal to $P_1/P_2=1.4$. The third harmonic was detected in an observation of a decaying transverse oscillation of coronal loops in \citet{Duckenfield2019A&A...632A..64D}, with the period ratio of the fundamental to the third harmonic equal to $P_1/P_3=2.61$. Considering our 1D eigenvalue problem for our full-loop system, the corresponding period ratios obtained are $P_1/P_2=1.03$ and $P_1/P_3=1.77$. These values seem to significantly deviate from the observations. However, if we assume that the third, sixth, and ninth harmonics were misidentified for the first, second, and third harmonics, we obtain ratios that match the observations: $P_\mathrm{1,false}/P_\mathrm{2,false}=1.76$ and $P_\mathrm{1,false}/P_\mathrm{3,false}=2.57$. Unsurprisingly by this point, if we consider the first three harmonics of the coronal loop system, the same ratios give $P_1/P_2=1.75$ and $P_1/P_3=2.57$, which are again a good match to the observations. Although a larger statistical sample is needed for observations and simulations of oscillating stratified loops, the period ratios above indicate that the observed modes for standing oscillations of longer coronal loops either are purely coronal in nature, with the part of the loop below the corona not being perturbed, or are the modes that exhibit nodes near the base of the corona. 

Our analysis so far has focused only on our loop setup, which models a longer coronal loop with length $200$ Mm. However, recent observations of short coronal and transition region loops with length $\leq 50$ Mm have also revealed transverse decayless oscillations. Unlike with the longer coronal loops, most studies of decayless oscillations in short loops \citep[e.g.][]{GaoYuhang2022ApJ...930...55G, Petrova2023ApJ...946...36P, ShrivastavArpitKumar2024A&A...685A..36S, GaoYuhang2024A&A...681L...4G} could not conclusively characterise these waves as standing or propagating, leaving their interpretation still open. Despite that, the assumption of them being standing waves has been used extensively in these studies when considering them for seismology.  \citet{Lim2024A&A...690L...8L} show that under-sampling can lead to overestimation of the oscillation periods, producing values that match the observations for shorter loops. The difference with respect to the longer coronal loops is that a higher percentage of the total length of these short loops will be located below the corona. Therefore, we expect these short loops to exhibit qualitatively different spatial profiles of their kink oscillation modes. To get better insight into the oscillatory profiles of these shorter loops, we repeated our analysis with the 1D wave equation eigenvalue system for a short loop of length $30$ Mm. The kink speed profile that we used as phase speed in Eq. \ref{eq:1D} was taken from the numerical model of a short coronal loop model that was used in \citet{Gao2023ApJ...955...73G}. This profile of the kink speed is shown in the left panel of Fig. \ref{fig:modesYuhang1D}. The approximate height of the top of the transition region is this model is located at height $z=4.2$ Mm above each footpoint. As a reference, the coronal part of the kink speed profile is also shown. The right panel of Fig. \ref{fig:modesYuhang1D} shows the spatial profiles of the standing kink modes for this system and their respective frequencies. For reference, the derived frequencies of the first, second and third harmonics reported in \citet{Gao2023ApJ...955...73G} are $f_1=13.98$ mHz, $f_2=18.62$ mHz and $f_3=24.04$ mHz, with $P_1/P_2=1.33$ and $P_1/P_3=1.72$. From this panel, we see that it is easier to distinguish between the first and third harmonics, due to the proximity of the loop apex to the base of the corona. In fact, the oscillation of such a low-lying loop would likely be visible in transition region lines, as was the case with the observed decayless oscillations in transition region loops reported in \citet{GaoYuhang2024A&A...681L...4G}. Therefore, in the case of shorter, low-lying loops we expect less ambiguity over the order of the observed standing kink modes, than in longer coronal loops.

The aforementioned possible ambiguity over the order of standing kink modes in coronal loops, as it translates to the observations of transverse decaying and decayless oscillations can have seismological implications. For example, the observed periods of the fundamental mode and the loop length can be used to calculate the kink speed and through that average magnetic field, when combined with estimates of the plasma density and the ratio of densities inside and outside of the oscillating loops \citep[see for example][]{anfinogentov2019ApJ,ZhongLongLoop2023NatSR..1312963Z,GaoYuhang2024A&A...681L...4G}. In Fig. \ref{fig:decayless} we plot the oscillation period and corresponding kink speed with respect to the loop length, for data taken from a number of past studies \citep{wang2012,nistico2013,anfinogentov2013,anfinogentov2015,anfinogentov2019ApJ,ZhongSihui2022MNRAS.513.1834Z,ZhongSihui2022MNRAS.516.5989Z,GaoYuhang2022ApJ...930...55G,GaoYuhang2024A&A...681L...4G,LiandLong2023ApJ...944....8L,Petrova2023ApJ...946...36P,ShrivastavArpitKumar2024A&A...685A..36S}. The left panel shows the scatter plot of the period versus the loop length. The blue circles represent the original observations for short loops, with a length of less than $50$ Mm (dashed vertical line), the value of which was arbitrary chosen. We show the original observations of long coronal loops and the best linear fit ($P=a\,L$) with coefficient $a=1.103\pm 0.043$ s Mm$^{-1}$. Estimating the kink speed as $C_\mathrm{Kink}=2L/P_\mathrm{Kink}$, where $P_\mathrm{Kink}$ is considered as the period of the fundamental standing kink mode, we get an average kink speed $C_\mathrm{Kink}=1.81$ Mm s$^{-1}$. The estimated kink speed with respect to the loop length for each observation is shown in the right panel of Fig. \ref{fig:decayless}. Considering now the possible ambiguity between the first and the third kink modes in long loops, we corrected the values of the observed periods in Fig. \ref{fig:decayless} by a factor of $1.77$, since $P_1=1.77\,P_3$. Here we used the results for our system as a suggested average correction for the observations of the long loops ($L>50$ Mm). The corrected values for the period and the estimated kink speed are shown in both panels. From the new linear fit we get a coefficient $a=1.953\pm0.076$ and a new average kink speed of $C_\mathrm{Kink}=1.02$ Mm s$^{-1}$. Using common seismology techniques \citep[e.g.][]{anfinogentov2019ApJ,ZhongLongLoop2023NatSR..1312963Z,GaoYuhang2024A&A...681L...4G}, we can calculate the average magnetic field of an oscillating loop with the following formula:
\begin{equation}
    B = C_\mathrm{Kink} \sqrt{\frac{\mu_0(\rho_i+\rho_e)}{2}} = \frac{2L}{P_\mathrm{Kink}} \sqrt{\frac{\mu_0(\rho_i+\rho_e)}{2}},
\end{equation}
where $\mu_0$ is the magnetic permeability. If we take for the observed $P_\mathrm{Kink} = P_3$ instead of $P_1$, due to the described ambiguity, then for the average calculated magnetic field we have 
\begin{equation}
    B_\mathrm{calc} = \frac{2L}{P_3} \sqrt{\frac{\mu_0(\rho_i+\rho_e)}{2}} = \frac{P_1}{P_3} \left(\frac{2L}{P_1} \sqrt{\frac{\mu_0(\rho_i+\rho_e)}{2}} \right)
,\end{equation}
and therefore $B_\mathrm{calc} = (P_1/P_3)\, B_\mathrm{real}$. For a mean background magnetic field of $B_\mathrm{real}=30$ G, as in our model, the field calculated from the observed quantities will be overestimated by a factor of $ P_1/ P_3 $ (here $1.77$), giving $B_\mathrm{calc} = 1.77\times 30 \sim 53$ G.
\section{Discussion and conclusions} \label{sec:discussions}

We explored the nature of the spectra in a gravitationally stratified loop with footpoints anchored in chromospheric plasma by performing transverse oscillations. Our base model consists of a 3D straight flux tube, which is perturbed in three different simulations by (a) a footpoint linearly polarised broadband velocity driver, (b) a symmetric sinusoidal velocity pulse, and (c) an asymmetric off-centre Gaussian pulse. We limited our 3D study mostly to the coronal part of our system and treated the chromosphere as a mass reservoir of near-constant temperature. We also performed a complimentary analysis, solving the generalised eigenvalue problem for the 1D wave equation with nodes at the footpoints, for a system with phase speed equal to the kink speed profile of the 3D loop. Finally, we performed forward modelling with our simulation data and then explored the seismological implications of our findings.

When tracking the centre of mass of our loop, the oscillation spectra reveal the existence of multiple harmonics when either the broadband driver or the off-centre velocity pulse are used to perturb the flux tube, as shown in Fig. \ref{fig:cmspectra}. Our 1D analysis reveals that the observed signatures correspond to the frequencies of the system eigenmodes (Fig. \ref{fig:modes1D}), which have been deformed due to the large gradients in the values of density with height. The so-called half-harmonic that was reported in our past studies \citep{Karampelas2024A&A...681L...6K,Karampelas2024A&A...688A..80K} is now understood to consist of the deformed fundamental kink mode and the second harmonic, and in the case of the broadband driver it also has the signatures of driving frequencies. What was understood in past studies as the fundamental mode is in fact the third harmonic of the system. However, our analysis shows that this third harmonic of the full-length loop (and also the sixth, ninth, and twelfth harmonics) exhibits similar frequencies and spatial profiles in the corona as the fundamental mode (and also the second, third, and fourth harmonics) of the coronal part of that loop. This similarity is due to the nodes that these harmonics exhibit near the base of the corona for our system. 

Creating synthetic observations of our 3D simulations and tracking the centre of the loop at different resolutions, we extracted the oscillation frequencies that would have been detected in real observations of a system matching our model. The broadband driver excites a standing wave with many frequencies present, consisting of multiple harmonics and frequencies associated with the driving. Most observations of decayless oscillations seem to have a single dominant period \citep[e.g.][]{anfinogentov2015,Petrova2023ApJ...946...36P}, and it is rare to detect higher-order harmonics \citep{duckenfield2018ApJ}. As such, our findings cast doubts on the idea that broadband drivers could be one of the possible mechanisms responsible for the excitation of decayless oscillations in coronal loops. For a conclusive answer on the nature of the drivers of these oscillations, realistic driving profiles that follow observational data need to be employed in future numerical studies.

Focusing on the oscillations excited by the initial velocity pulses, our spectra from the synthetic observations in the 171 \r{A} line reveal only one dominant frequency band, which matches the third harmonic of the full-length loop. This mode, which also resembles the fundamental mode of the coronal part of the loop, can easily be misidentified as the fundamental mode of the full-length loop. The time-distance maps of the oscillating loops at the apex in Fig. \ref{fig:tdmaps} show signals of decaying amplitude over time with a single frequency. However, the oscillations get progressively harder to track after the first two to three periods due to the small initial amplitude. Considering the overlap of features in real observations, or the different levels of contrast of the intensity between the loop and the background, it is possible for these decaying oscillations to be categorised as decayless with a fundamental frequency at $\sim 4.75$ mHz  unless a longer part of the signal is clearly visible for a longer time. It is worth noting that some recent observations of decayless oscillations have shown them to have relatively short durations \citep[up to four cycles; e.g.][]{ShrivastavArpitKumar2024A&A...685A..36S, Lim2024A&A...689A..16L}, creating ambiguity as to their interpretation as decaying or decayless oscillations, similar to what was discussed for our data. 

As mentioned, the third harmonic of our oscillating loop can be misidentified as the fundamental kink mode of the loop \citep[see also][]{HowsonBrue2023MNRAS.526..499H}. This misidentification is caused by the deformation of the harmonic modes due to the low values of the kink speed and the density scale height at the loop footpoints, below the transition region. However, our 3D numerical setup is characterised by a very wide chromosphere ($\Delta z = 5$\,Mm) and an unphysically broadened transition region, which are not representative of the solar atmosphere. Therefore, our results are expected to differ from those for a more realistic profile of the solar atmosphere. Past studies have treated coronal loops as resonant cavities for low-frequency fluctuations \citep[e.g.][]{Hollweg1984SoPh...91..269H, Verdini2012A&A...538A..70V}, with resonant waves being confined in the coronal part and experiencing wave leakage through the sharp transition region. It is unclear, therefore, how a realistic chromospheric and transition region profile will affect the kink modes of our 3D model. Our 1D analysis predicts the existence of the previously mentioned deformed harmonics, as well as the respective values for the frequencies of these modes, as derived from the 3D simulations. However, as stated in Sect. \ref{sec:1D}, the thin tube approximation used to derive Eq. \ref{eq:1D} is not valid below the transition region for our model. This means that the results of our 1D analysis can only be taken as an approximation of the actual transverse modes of the system. In addition, the results of the 1D analysis also depend on the profiles of the loop and the lower solar atmosphere considered in our model. In Appendix \ref{sec:appendix} we show the effects of different kink speed profiles on the solutions of Eq. \ref{eq:1D}, highlighting this further. From the above, it becomes clear that a proper parameter study based on realistic profiles of the lower solar atmosphere is required (a) to enable a detailed examination of the spatial profiles of the oscillation modes in the 3D simulations, (b) to examine the validity of using Eq. \ref{eq:1D} for loops anchored in chromospheric plasma, and (c) to conclude on the possibility of misidentifying the order of the observed transverse oscillations in loops, and the applicability of our findings.

The presence of transverse oscillations has been confirmed in shorter loops \citep[e.g.][]{GaoYuhang2022ApJ...930...55G, LiandLong2023ApJ...944....8L, Petrova2023ApJ...946...36P, ShrivastavArpitKumar2024A&A...685A..36S,GaoYuhang2024A&A...681L...4G}, with the recent study by \citet{Lim2024A&A...690L...8L} showing that they can be interpreted as standing waves with periods shorter than the temporal resolution of our current observatories. This suggests that the transverse oscillations observed in coronal loops might not necessarily be confined to the corona, similar to how the kink modes in our 3D numerical model also persist along the entire length of the loop. This seems to invalidate the treatment of loops as resonant cavities that confine waves in their coronal part while taking the effects of leakage from the footpoints into account  \citep[e.g.][]{Hollweg1984SoPh...91..269H, Verdini2012A&A...538A..70V}. However, we again note that our models are characterised by an artificially broadened transition region. The effects of a sharp transition region on transverse standing waves in loops need to be properly explored in future numerical studies. Due to this uncertainty, observations at chromospheric and transition region lines at the footpoints of oscillating loops are necessary to identify the observed kink modes as either purely coronal modes or modes of the full-loop system before making any seismological estimations.

More advanced techniques of spatial coronal seismology, such as the $\chi^2$ optimisation process shown in this study, produce results with a good agreement between full-loop harmonics and coronal loop harmonics. However, many data points along the entire length of the loop are required to properly constrain the estimated mean magnetic field values. Therefore, we conclude that the current approach in seismology, where only the standing modes in the coronal part of the loops are considered, can only be justified if there is definitive proof that the oscillatory motion is confined to the corona. Unveiling the dynamics of oscillating loops in the lower corona, the chromosphere and the transition region through simultaneous high-resolution and high-cadence observations from upcoming missions such as MUSE is essential for future seismological studies. 

\begin{acknowledgements}
K.K. acknowledges support by an FWO (Fonds voor Wetenschappelijk Onderzoek – Vlaanderen) postdoctoral fellowship (1273221N). DL was supported by a Senior Research Project (G088021N) of the FWO Vlaanderen. TVD was supported by the C1 grant TRACEspace of Internal Funds KU Leuven, and a Senior Research Project (G088021N) of the FWO Vlaanderen. Furthermore, TVD received financial support from the Flemish Government under the long-term structural Methusalem funding program, project SOUL: Stellar evolution in full glory, grant METH/24/012 at KU Leuven. The research that led to these results was subsidised by the Belgian Federal Science Policy Office through the contract B2/223/P1/CLOSE-UP. It is also part of the DynaSun project and has thus received funding under the Horizon Europe programme of the European Union under grant agreement (no. 101131534). Views and opinions expressed are however those of the author(s) only and do not necessarily reflect those of the European Union and therefore the European Union cannot be held responsible for them. Y.G. was supported by China Scholarship Council under file No. 202206010018. The computational resources and services used in this work were provided by the VSC (Flemish Supercomputer Center), funded by the Research Foundation Flanders (FWO) and the Flemish Government – department EWI. 
\end{acknowledgements}

\bibliographystyle{aa}
\bibliography{paper}

\begin{thebibliography}{81}
\expandafter\ifx\csname natexlab\endcsname\relax\def\natexlab#1{#1}\fi

\bibitem[{{Afanasyev} {et~al.}(2019){Afanasyev}, {Karampelas}, \& {Van Doorsselaere}}]{afanasev2019}
{Afanasyev}, A., {Karampelas}, K., \& {Van Doorsselaere}, T. 2019, \apj, 876, 100

\bibitem[{{Afanasyev} {et~al.}(2020){Afanasyev}, {Van Doorsselaere}, \& {Nakariakov}}]{afanasyev2020decayless}
{Afanasyev}, A.~N., {Van Doorsselaere}, T., \& {Nakariakov}, V.~M. 2020, \aap, 633, L8

\bibitem[{{Andries} {et~al.}(2005{\natexlab{a}}){Andries}, {Arregui}, \& {Goossens}}]{Andries2005ApJ...624L..57A}
{Andries}, J., {Arregui}, I., \& {Goossens}, M. 2005{\natexlab{a}}, \apjl, 624, L57

\bibitem[{{Andries} {et~al.}(2005{\natexlab{b}}){Andries}, {Goossens}, {Hollweg}, {Arregui}, \& {Van Doorsselaere}}]{andries2005}
{Andries}, J., {Goossens}, M., {Hollweg}, J.~V., {Arregui}, I., \& {Van Doorsselaere}, T. 2005{\natexlab{b}}, \aap, 430, 1109

\bibitem[{{Andries} {et~al.}(2009){Andries}, {van Doorsselaere}, {Roberts}, {Verth}, {Verwichte}, \& {Erd{\'e}lyi}}]{Andries2009SSRv..149....3A}
{Andries}, J., {van Doorsselaere}, T., {Roberts}, B., {et~al.} 2009, \ssr, 149, 3

\bibitem[{{Anfinogentov} {et~al.}(2013){Anfinogentov}, {Nistic{\`o}}, \& {Nakariakov}}]{anfinogentov2013}
{Anfinogentov}, S., {Nistic{\`o}}, G., \& {Nakariakov}, V.~M. 2013, \aap, 560, A107

\bibitem[{{Anfinogentov} \& {Nakariakov}(2019)}]{anfinogentov2019ApJ}
{Anfinogentov}, S.~A. \& {Nakariakov}, V.~M. 2019, \apjl, 884, L40

\bibitem[{{Anfinogentov} {et~al.}(2015){Anfinogentov}, {Nakariakov}, \& {Nistic{\`o}}}]{anfinogentov2015}
{Anfinogentov}, S.~A., {Nakariakov}, V.~M., \& {Nistic{\`o}}, G. 2015, \aap, 583, A136

\bibitem[{Antolin {et~al.}(2016)Antolin, Moortel, Doorsselaere, \& Yokoyama}]{antolin2016}
Antolin, P., Moortel, I.~D., Doorsselaere, T.~V., \& Yokoyama, T. 2016, The Astrophysical Journal Letters, 830, L22

\bibitem[{{Aschwanden} {et~al.}(1999){Aschwanden}, {Fletcher}, {Schrijver}, \& {Alexander}}]{aschwanden1999}
{Aschwanden}, M.~J., {Fletcher}, L., {Schrijver}, C.~J., \& {Alexander}, D. 1999, \apj, 520, 880

\bibitem[{{Aschwanden} \& {Schrijver}(2011)}]{AschwandenSchrijver2011ApJ...736..102A}
{Aschwanden}, M.~J. \& {Schrijver}, C.~J. 2011, \apj, 736, 102

\bibitem[{{Chen} {et~al.}(2023){Chen}, {Guo}, {Ding}, \& {Erd{\'e}lyi}}]{ChenErdelyi2023A&A...678A.205C}
{Chen}, G.~Y., {Guo}, Y., {Ding}, M.~D., \& {Erd{\'e}lyi}, R. 2023, \aap, 678, A205

\bibitem[{{Cheung} {et~al.}(2022){Cheung}, {Mart{\'\i}nez-Sykora}, {Testa}, {De Pontieu}, {Chintzoglou}, {Rempel}, {Polito}, {Kerr}, {Reeves}, {Fletcher}, {Jin}, {N{\'o}brega-Siverio}, {Danilovic}, {Antolin}, {Allred}, {Hansteen}, {Ugarte-Urra}, {DeLuca}, {Longcope}, {Takasao}, {DeRosa}, {Boerner}, {Jaeggli}, {Nitta}, {Daw}, {Carlsson}, {Golub}, \& {The}}]{Cheung2022ApJ...926...53C}
{Cheung}, M. C.~M., {Mart{\'\i}nez-Sykora}, J., {Testa}, P., {et~al.} 2022, \apj, 926, 53

\bibitem[{{Chitta} {et~al.}(2012){Chitta}, {van Ballegooijen}, {Rouppe van der Voort}, {DeLuca}, \& {Kariyappa}}]{Chitta2012ApJ...752...48C}
{Chitta}, L.~P., {van Ballegooijen}, A.~A., {Rouppe van der Voort}, L., {DeLuca}, E.~E., \& {Kariyappa}, R. 2012, \apj, 752, 48

\bibitem[{{De Moortel} \& {Brady}(2007)}]{DeMoortel2007ApJ...664.1210D}
{De Moortel}, I. \& {Brady}, C.~S. 2007, \apj, 664, 1210

\bibitem[{{De Moortel} \& {Howson}(2022)}]{DeMoortel2022ApJ...941...85D}
{De Moortel}, I. \& {Howson}, T.~A. 2022, \apj, 941, 85

\bibitem[{{De Pontieu} {et~al.}(2022){De Pontieu}, {Testa}, {Mart{\'\i}nez-Sykora}, {Antolin}, {Karampelas}, {Hansteen}, {Rempel}, {Cheung}, {Reale}, {Danilovic}, {Pagano}, {Polito}, {De Moortel}, {N{\'o}brega-Siverio}, {Van Doorsselaere}, {Petralia}, {Asgari-Targhi}, {Boerner}, {Carlsson}, {Chintzoglou}, {Daw}, {DeLuca}, {Golub}, {Matsumoto}, {Ugarte-Urra}, {McIntosh}, \& {the MUSE Team}}]{depontieu2022ApJ...926...52D...MUSE}
{De Pontieu}, B., {Testa}, P., {Mart{\'\i}nez-Sykora}, J., {et~al.} 2022, \apj, 926, 52

\bibitem[{{Duckenfield} {et~al.}(2018){Duckenfield}, {Anfinogentov}, {Pascoe}, \& {Nakariakov}}]{duckenfield2018ApJ}
{Duckenfield}, T., {Anfinogentov}, S.~A., {Pascoe}, D.~J., \& {Nakariakov}, V.~M. 2018, \apjl, 854, L5

\bibitem[{{Duckenfield} {et~al.}(2019){Duckenfield}, {Goddard}, {Pascoe}, \& {Nakariakov}}]{Duckenfield2019A&A...632A..64D}
{Duckenfield}, T.~J., {Goddard}, C.~R., {Pascoe}, D.~J., \& {Nakariakov}, V.~M. 2019, \aap, 632, A64

\bibitem[{{Dymova} \& {Ruderman}(2005)}]{DymovaRuderman2005SoPh..229...79D}
{Dymova}, M.~V. \& {Ruderman}, M.~S. 2005, \solphys, 229, 79

\bibitem[{{Edwin} \& {Roberts}(1983)}]{edwin1983wave}
{Edwin}, P.~M. \& {Roberts}, B. 1983, \solphys, 88, 179

\bibitem[{{Erd{\'e}lyi} \& {Verth}(2007)}]{ErdelyiVerth2007A&A...462..743E}
{Erd{\'e}lyi}, R. \& {Verth}, G. 2007, \aap, 462, 743

\bibitem[{{Gao} {et~al.}(2023){Gao}, {Guo}, {Van Doorsselaere}, {Tian}, \& {Skirvin}}]{Gao2023ApJ...955...73G}
{Gao}, Y., {Guo}, M., {Van Doorsselaere}, T., {Tian}, H., \& {Skirvin}, S.~J. 2023, \apj, 955, 73

\bibitem[{{Gao} {et~al.}(2024){Gao}, {Hou}, {Van Doorsselaere}, \& {Guo}}]{GaoYuhang2024A&A...681L...4G}
{Gao}, Y., {Hou}, Z., {Van Doorsselaere}, T., \& {Guo}, M. 2024, \aap, 681, L4

\bibitem[{{Gao} {et~al.}(2022){Gao}, {Tian}, {Van Doorsselaere}, \& {Chen}}]{GaoYuhang2022ApJ...930...55G}
{Gao}, Y., {Tian}, H., {Van Doorsselaere}, T., \& {Chen}, Y. 2022, \apj, 930, 55

\bibitem[{{Guo} {et~al.}(2023){Guo}, {Duckenfield}, {Van Doorsselaere}, {Karampelas}, {Pelouze}, \& {Gao}}]{mingzhe2023ApJ...949L...1G}
{Guo}, M., {Duckenfield}, T., {Van Doorsselaere}, T., {et~al.} 2023, \apjl, 949, L1

\bibitem[{{Handy} {et~al.}(1999){Handy}, {Acton}, {Kankelborg}, {Wolfson}, {Akin}, {Bruner}, {Caravalho}, {Catura}, {Chevalier}, {Duncan}, {Edwards}, {Feinstein}, {Freeland}, {Friedlaender}, {Hoffmann}, {Hurlburt}, {Jurcevich}, {Katz}, {Kelly}, {Lemen}, {Levay}, {Lindgren}, {Mathur}, {Meyer}, {Morrison}, {Morrison}, {Nightingale}, {Pope}, {Rehse}, {Schrijver}, {Shine}, {Shing}, {Strong}, {Tarbell}, {Title}, {Torgerson}, {Golub}, {Bookbinder}, {Caldwell}, {Cheimets}, {Davis}, {Deluca}, {McMullen}, {Warren}, {Amato}, {Fisher}, {Maldonado}, \& {Parkinson}}]{Handy1999SoPh..187..229H}
{Handy}, B.~N., {Acton}, L.~W., {Kankelborg}, C.~C., {et~al.} 1999, \solphys, 187, 229

\bibitem[{{Hollweg}(1984)}]{Hollweg1984SoPh...91..269H}
{Hollweg}, J.~V. 1984, \solphys, 91, 269

\bibitem[{{Howson} \& {De Moortel}(2023)}]{Howson2023Physi...5..140H}
{Howson}, T. \& {De Moortel}, I. 2023, Physics, 5, 140

\bibitem[{{Howson} \& {Breu}(2023)}]{HowsonBrue2023MNRAS.526..499H}
{Howson}, T.~A. \& {Breu}, C. 2023, \mnras, 526, 499

\bibitem[{{Howson} \& {De Moortel}(2022)}]{Howson2022A&A...661A.144H}
{Howson}, T.~A. \& {De Moortel}, I. 2022, \aap, 661, A144

\bibitem[{{Karampelas} \& {Van Doorsselaere}(2020)}]{karampelas2020ApJ}
{Karampelas}, K. \& {Van Doorsselaere}, T. 2020, \apjl, 897, L35

\bibitem[{{Karampelas} \& {Van Doorsselaere}(2021)}]{karampelas2021ApJ...908L...7K}
{Karampelas}, K. \& {Van Doorsselaere}, T. 2021, \apjl, 908, L7

\bibitem[{{Karampelas} \& {Van Doorsselaere}(2024)}]{Karampelas2024A&A...681L...6K}
{Karampelas}, K. \& {Van Doorsselaere}, T. 2024, \aap, 681, L6

\bibitem[{{Karampelas} {et~al.}(2024){Karampelas}, {Van Doorsselaere}, {Guo}, {Duckenfield}, \& {Pelouze}}]{Karampelas2024A&A...688A..80K}
{Karampelas}, K., {Van Doorsselaere}, T., {Guo}, M., {Duckenfield}, T., \& {Pelouze}, G. 2024, \aap, 688, A80

\bibitem[{{Karampelas} {et~al.}(2019){Karampelas}, {Van Doorsselaere}, {Pascoe}, {Guo}, \& {Antolin}}]{karampelas2019amp}
{Karampelas}, K., {Van Doorsselaere}, T., {Pascoe}, J.~D., {Guo}, M., \& {Antolin}, P. 2019, Frontiers in Astronomy and Space Sciences, 6, 38

\bibitem[{{Kohutova} \& {Verwichte}(2017)}]{KohutovaVerwichte2017A&A...606A.120K}
{Kohutova}, P. \& {Verwichte}, E. 2017, \aap, 606, A120

\bibitem[{{Kohutova} \& {Verwichte}(2018)}]{KohutovaVerwichte2018A&A...613L...3K}
{Kohutova}, P. \& {Verwichte}, E. 2018, \aap, 613, L3

\bibitem[{{Lemen} {et~al.}(2012){Lemen}, {Title}, {Akin}, {Boerner}, {Chou}, {Drake}, {Duncan}, {Edwards}, {Friedlaender}, {Heyman}, {Hurlburt}, {Katz}, {Kushner}, {Levay}, {Lindgren}, {Mathur}, {McFeaters}, {Mitchell}, {Rehse}, {Schrijver}, {Springer}, {Stern}, {Tarbell}, {Wuelser}, {Wolfson}, {Yanari}, {Bookbinder}, {Cheimets}, {Caldwell}, {Deluca}, {Gates}, {Golub}, {Park}, {Podgorski}, {Bush}, {Scherrer}, {Gummin}, {Smith}, {Auker}, {Jerram}, {Pool}, {Soufli}, {Windt}, {Beardsley}, {Clapp}, {Lang}, \& {Waltham}}]{Lemen2012SoPh..275...17L}
{Lemen}, J.~R., {Title}, A.~M., {Akin}, D.~J., {et~al.} 2012, \solphys, 275, 17

\bibitem[{{Li} \& {Long}(2023)}]{LiandLong2023ApJ...944....8L}
{Li}, D. \& {Long}, D.~M. 2023, \apj, 944, 8

\bibitem[{{Lim} {et~al.}(2023){Lim}, {Van Doorsselaere}, {Berghmans}, {Morton}, {Pant}, \& {Mandal}}]{Lim2023ApJ...952L..15L}
{Lim}, D., {Van Doorsselaere}, T., {Berghmans}, D., {et~al.} 2023, \apjl, 952, L15

\bibitem[{{Lim} {et~al.}(2024{\natexlab{a}}){Lim}, {Van Doorsselaere}, {Berghmans}, \& {Petrova}}]{Lim2024A&A...689A..16L}
{Lim}, D., {Van Doorsselaere}, T., {Berghmans}, D., \& {Petrova}, E. 2024{\natexlab{a}}, \aap, 689, A16

\bibitem[{{Lim} {et~al.}(2024{\natexlab{b}}){Lim}, {Van Doorsselaere}, {Nakariakov}, {Kolotkov}, {Gao}, \& {Berghmans}}]{Lim2024A&A...690L...8L}
{Lim}, D., {Van Doorsselaere}, T., {Nakariakov}, V.~M., {et~al.} 2024{\natexlab{b}}, \aap, 690, L8

\bibitem[{Linker {et~al.}(2001)Linker, Lionello, Miki{\'c}, \& Amari}]{LinkerEtAl2001}
Linker, J.~A., Lionello, R., Miki{\'c}, Z., \& Amari, T. 2001, \jgr, 106, 25165

\bibitem[{Lionello {et~al.}(2009)Lionello, Linker, \& Miki{\'c}}]{LionelloEtAl2009}
Lionello, R., Linker, J.~A., \& Miki{\'c}, Z. 2009, \apj, 690, 902

\bibitem[{{Mignone} {et~al.}(2007){Mignone}, {Bodo}, {Massaglia}, {Matsakos}, {Tesileanu}, {Zanni}, \& {Ferrari}}]{mignonePLUTO2007}
{Mignone}, A., {Bodo}, G., {Massaglia}, S., {et~al.} 2007, \apjs, 170, 228

\bibitem[{Miki{\'c} {et~al.}(2013)Miki{\'c}, Lionello, Mok, Linker, \& Winebarger}]{MikicEtAl2013}
Miki{\'c}, Z., Lionello, R., Mok, Y., Linker, J.~A., \& Winebarger, A.~R. 2013, \apj, 773, 94

\bibitem[{{Morton} {et~al.}(2023){Morton}, {Sharma}, {Tajfirouze}, \& {Miriyala}}]{Morton2023RvMPP...7...17M}
{Morton}, R.~J., {Sharma}, R., {Tajfirouze}, E., \& {Miriyala}, H. 2023, Reviews of Modern Plasma Physics, 7, 17

\bibitem[{{Morton} {et~al.}(2013){Morton}, {Verth}, {Fedun}, {Shelyag}, \& {Erd{\'e}lyi}}]{Morton2013ApJ...768...17M}
{Morton}, R.~J., {Verth}, G., {Fedun}, V., {Shelyag}, S., \& {Erd{\'e}lyi}, R. 2013, \apj, 768, 17

\bibitem[{Nakariakov {et~al.}(2021)Nakariakov, Anfinogentov, Antolin, Jain, Kolotkov, Kupriyanova, Li, Magyar, Nistico, Pascoe, Srivastava, Terradas, Farahani, Verth, Yuan, \& Zimovets}]{NakariakovEtAl2021}
Nakariakov, V.~M., Anfinogentov, S.~A., Antolin, P., {et~al.} 2021, \ssr, 217, 73

\bibitem[{{Nakariakov} {et~al.}(2016){Nakariakov}, {Anfinogentov}, {Nistic{\`o}}, \& {Lee}}]{nakariakov2016}
{Nakariakov}, V.~M., {Anfinogentov}, S.~A., {Nistic{\`o}}, G., \& {Lee}, D.-H. 2016, \aap, 591, L5

\bibitem[{{Nakariakov} {et~al.}(2009){Nakariakov}, {Aschwanden}, \& {van Doorsselaere}}]{nakariakov2009}
{Nakariakov}, V.~M., {Aschwanden}, M.~J., \& {van Doorsselaere}, T. 2009, \aap, 502, 661

\bibitem[{{Nakariakov} {et~al.}(1999){Nakariakov}, {Ofman}, {Deluca}, {Roberts}, \& {Davila}}]{nakariakov1999}
{Nakariakov}, V.~M., {Ofman}, L., {Deluca}, E.~E., {Roberts}, B., \& {Davila}, J.~M. 1999, Science, 285, 862

\bibitem[{{Nakariakov} \& {Verwichte}(2005)}]{Nakariakov2005LRSP}
{Nakariakov}, V.~M. \& {Verwichte}, E. 2005, Living Reviews in Solar Physics, 2, 3

\bibitem[{{Nechaeva} {et~al.}(2019){Nechaeva}, {Zimovets}, {Nakariakov}, \& {Goddard}}]{Nechaeva2019ApJS}
{Nechaeva}, A., {Zimovets}, I.~V., {Nakariakov}, V.~M., \& {Goddard}, C.~R. 2019, \apjs, 241, 31

\bibitem[{{Nistic{\`o}} {et~al.}(2013){Nistic{\`o}}, {Nakariakov}, \& {Verwichte}}]{nistico2013}
{Nistic{\`o}}, G., {Nakariakov}, V.~M., \& {Verwichte}, E. 2013, \aap, 552, A57

\bibitem[{{Orlando} {et~al.}(2008){Orlando}, {Bocchino}, {Reale}, {Peres}, \& {Pagano}}]{Orlando2008ApJ}
{Orlando}, S., {Bocchino}, F., {Reale}, F., {Peres}, G., \& {Pagano}, P. 2008, \apj, 678, 274

\bibitem[{{Pelouze} {et~al.}(2023){Pelouze}, {Van Doorsselaere}, {Karampelas}, {Riedl}, \& {Duckenfield}}]{pelouze2023A&A...672A.105P}
{Pelouze}, G., {Van Doorsselaere}, T., {Karampelas}, K., {Riedl}, J.~M., \& {Duckenfield}, T. 2023, \aap, 672, A105

\bibitem[{{Petrova} {et~al.}(2023){Petrova}, {Magyar}, {Van Doorsselaere}, \& {Berghmans}}]{Petrova2023ApJ...946...36P}
{Petrova}, E., {Magyar}, N., {Van Doorsselaere}, T., \& {Berghmans}, D. 2023, \apj, 946, 36

\bibitem[{{Roberts} {et~al.}(1984){Roberts}, {Edwin}, \& {Benz}}]{Roberts1984ApJ...279..857R}
{Roberts}, B., {Edwin}, P.~M., \& {Benz}, A.~O. 1984, \apj, 279, 857

\bibitem[{{Rochus} {et~al.}(2020){Rochus}, {Auch{\`e}re}, {Berghmans}, {Harra}, {Schmutz}, {Sch{\"u}hle}, {Addison}, {Appourchaux}, {Aznar Cuadrado}, {Baker}, {Barbay}, {Bates}, {BenMoussa}, {Bergmann}, {Beurthe}, {Borgo}, {Bonte}, {Bouzit}, {Bradley}, {B{\"u}chel}, {Buchlin}, {B{\"u}chner}, {Cab{\'e}}, {Cadiergues}, {Chaigneau}, {Chares}, {Choque Cortez}, {Coker}, {Condamin}, {Coumar}, {Curdt}, {Cutler}, {Davies}, {Davison}, {Defise}, {Del Zanna}, {Delmotte}, {Delouille}, {Dolla}, {Dumesnil}, {D{\"u}rig}, {Enge}, {Fran{\c{c}}ois}, {Fourmond}, {Gillis}, {Giordanengo}, {Gissot}, {Green}, {Guerreiro}, {Guilbaud}, {Gyo}, {Haberreiter}, {Hafiz}, {Hailey}, {Halain}, {Hansotte}, {Hecquet}, {Heerlein}, {Hellin}, {Hemsley}, {Hermans}, {Hervier}, {Hochedez}, {Houbrechts}, {Ihsan}, {Jacques}, {J{\'e}r{\^o}me}, {Jones}, {Kahle}, {Kennedy}, {Klaproth}, {Kolleck}, {Koller}, {Kotsialos}, {Kraaikamp}, {Langer}, {Lawrenson}, {Le Clech'}, {Lenaerts}, {Liebecq}, {Linder}, {Long}, {Mampaey}, {Markiewicz-Innes}, {Marquet},
  {Marsch}, {Matthews}, {Mazy}, {Mazzoli}, {Meining}, {Meltchakov}, {Mercier}, {Meyer}, {Monecke}, {Monfort}, {Morinaud}, {Moron}, {Mountney}, {M{\"u}ller}, {Nicula}, {Parenti}, {Peter}, {Pfiffner}, {Philippon}, {Phillips}, {Plesseria}, {Pylyser}, {Rabecki}, {Ravet-Krill}, {Rebellato}, {Renotte}, {Rodriguez}, {Roose}, {Rosin}, {Rossi}, {Roth}, {Rouesnel}, {Roulliay}, {Rousseau}, {Ruane}, {Scanlan}, {Schlatter}, {Seaton}, {Silliman}, {Smit}, {Smith}, {Solanki}, {Spescha}, {Spencer}, {Stegen}, {Stockman}, {Szwec}, {Tamiatto}, {Tandy}, {Teriaca}, {Theobald}, {Tychon}, {van Driel-Gesztelyi}, {Verbeeck}, {Vial}, {Werner}, {West}, {Westwood}, {Wiegelmann}, {Willis}, {Winter}, {Zerr}, {Zhang}, \& {Zhukov}}]{Rochus2020A&A...642A...8R}
{Rochus}, P., {Auch{\`e}re}, F., {Berghmans}, D., {et~al.} 2020, \aap, 642, A8

\bibitem[{{Ruderman} \& {Petrukhin}(2021)}]{Ruderman2021MNRAS.501.3017R}
{Ruderman}, M.~S. \& {Petrukhin}, N.~S. 2021, \mnras, 501, 3017

\bibitem[{{Ruderman} {et~al.}(2021){Ruderman}, {Petrukhin}, \& {Pelinovsky}}]{Ruderman2021SoPh..296..124R}
{Ruderman}, M.~S., {Petrukhin}, N.~S., \& {Pelinovsky}, E. 2021, \solphys, 296, 124

\bibitem[{{Safari} {et~al.}(2007){Safari}, {Nasiri}, \& {Sobouti}}]{Safari2007}
{Safari}, H., {Nasiri}, S., \& {Sobouti}, Y. 2007, \aap, 470, 1111

\bibitem[{{Shi} {et~al.}(2021){Shi}, {Van Doorsselaere}, {Guo}, {Karampelas}, {Li}, \& {Antolin}}]{mijie2021ApJL}
{Shi}, M., {Van Doorsselaere}, T., {Guo}, M., {et~al.} 2021, \apj, 908, 233

\bibitem[{{Shrivastav} {et~al.}(2024){Shrivastav}, {Pant}, {Berghmans}, {Zhukov}, {Van Doorsselaere}, {Petrova}, {Banerjee}, {Lim}, \& {Verbeeck}}]{ShrivastavArpitKumar2024A&A...685A..36S}
{Shrivastav}, A.~K., {Pant}, V., {Berghmans}, D., {et~al.} 2024, \aap, 685, A36

\bibitem[{{Skirvin} {et~al.}(2023){Skirvin}, {Gao}, \& {Van Doorsselaere}}]{Skirvin2023ApJ...949...38S}
{Skirvin}, S.~J., {Gao}, Y., \& {Van Doorsselaere}, T. 2023, \apj, 949, 38

\bibitem[{{Sukarmadji} \& {Antolin}(2024)}]{Sukarmadji2024ApJ...961L..17S}
{Sukarmadji}, A. R.~C. \& {Antolin}, P. 2024, \apjl, 961, L17

\bibitem[{{Tian} {et~al.}(2012){Tian}, {McIntosh}, {Wang}, {Ofman}, {De Pontieu}, {Innes}, \& {Peter}}]{tian2012}
{Tian}, H., {McIntosh}, S.~W., {Wang}, T., {et~al.} 2012, \apj, 759, 144

\bibitem[{Van~Doorsselaere {et~al.}(2016)Van~Doorsselaere, Antolin, Yuan, Reznikova, \& Magyar}]{fomo2016}
Van~Doorsselaere, T., Antolin, P., Yuan, D., Reznikova, V., \& Magyar, N. 2016, Frontiers in Astronomy and Space Sciences, 3, 4

\bibitem[{Van~Doorsselaere {et~al.}(2008)Van~Doorsselaere, Nakariakov, \& Verwichte}]{tvd2008detection}
Van~Doorsselaere, T., Nakariakov, V.~M., \& Verwichte, E. 2008, The Astrophysical Journal Letters, 676, L73

\bibitem[{{Verdini} {et~al.}(2012){Verdini}, {Grappin}, \& {Velli}}]{Verdini2012A&A...538A..70V}
{Verdini}, A., {Grappin}, R., \& {Velli}, M. 2012, \aap, 538, A70

\bibitem[{{Verth} {et~al.}(2007){Verth}, {Van Doorsselaere}, {Erd{\'e}lyi}, \& {Goossens}}]{Verth2007A&A...475..341V}
{Verth}, G., {Van Doorsselaere}, T., {Erd{\'e}lyi}, R., \& {Goossens}, M. 2007, \aap, 475, 341

\bibitem[{{Verwichte} {et~al.}(2017){Verwichte}, {Antolin}, {Rowlands}, {Kohutova}, \& {Neukirch}}]{Verwichte2017A&A...598A..57V}
{Verwichte}, E., {Antolin}, P., {Rowlands}, G., {Kohutova}, P., \& {Neukirch}, T. 2017, \aap, 598, A57

\bibitem[{{Verwichte} \& {Kohutova}(2017)}]{VerwichteKohutova2017A&A...601L...2V}
{Verwichte}, E. \& {Kohutova}, P. 2017, \aap, 601, L2

\bibitem[{{Wang} {et~al.}(2012){Wang}, {Ofman}, {Davila}, \& {Su}}]{wang2012}
{Wang}, T., {Ofman}, L., {Davila}, J.~M., \& {Su}, Y. 2012, \apjl, 751, L27

\bibitem[{{Weberg} {et~al.}(2018){Weberg}, {Morton}, \& {McLaughlin}}]{Weberg2018ApJ...852...57W}
{Weberg}, M.~J., {Morton}, R.~J., \& {McLaughlin}, J.~A. 2018, \apj, 852, 57

\bibitem[{{Zhong} {et~al.}(2022{\natexlab{a}}){Zhong}, {Nakariakov}, {Kolotkov}, \& {Anfinogentov}}]{ZhongSihui2022MNRAS.513.1834Z}
{Zhong}, S., {Nakariakov}, V.~M., {Kolotkov}, D.~Y., \& {Anfinogentov}, S.~A. 2022{\natexlab{a}}, \mnras, 513, 1834

\bibitem[{{Zhong} {et~al.}(2023{\natexlab{a}}){Zhong}, {Nakariakov}, {Kolotkov}, {Chitta}, {Antolin}, {Verbeeck}, \& {Berghmans}}]{ZhongPolarisation2023NatCo..14.5298Z}
{Zhong}, S., {Nakariakov}, V.~M., {Kolotkov}, D.~Y., {et~al.} 2023{\natexlab{a}}, Nature Communications, 14, 5298

\bibitem[{{Zhong} {et~al.}(2022{\natexlab{b}}){Zhong}, {Nakariakov}, {Kolotkov}, {Verbeeck}, \& {Berghmans}}]{ZhongSihui2022MNRAS.516.5989Z}
{Zhong}, S., {Nakariakov}, V.~M., {Kolotkov}, D.~Y., {Verbeeck}, C., \& {Berghmans}, D. 2022{\natexlab{b}}, \mnras, 516, 5989

\bibitem[{{Zhong} {et~al.}(2023{\natexlab{b}}){Zhong}, {Nakariakov}, {Miao}, {Fu}, \& {Yuan}}]{ZhongLongLoop2023NatSR..1312963Z}
{Zhong}, S., {Nakariakov}, V.~M., {Miao}, Y., {Fu}, L., \& {Yuan}, D. 2023{\natexlab{b}}, Scientific Reports, 13, 12963

\end{thebibliography}

\begin{appendix}

\section{Solving the 1D wave equation eigenvalue problem for different phase speeds} \label{sec:appendix}
\begin{figure}[h!]
    \resizebox{\hsize}{!}{
    \includegraphics[trim={0.5cm 1.cm 0.4cm 0.3cm},clip,scale=0.25]{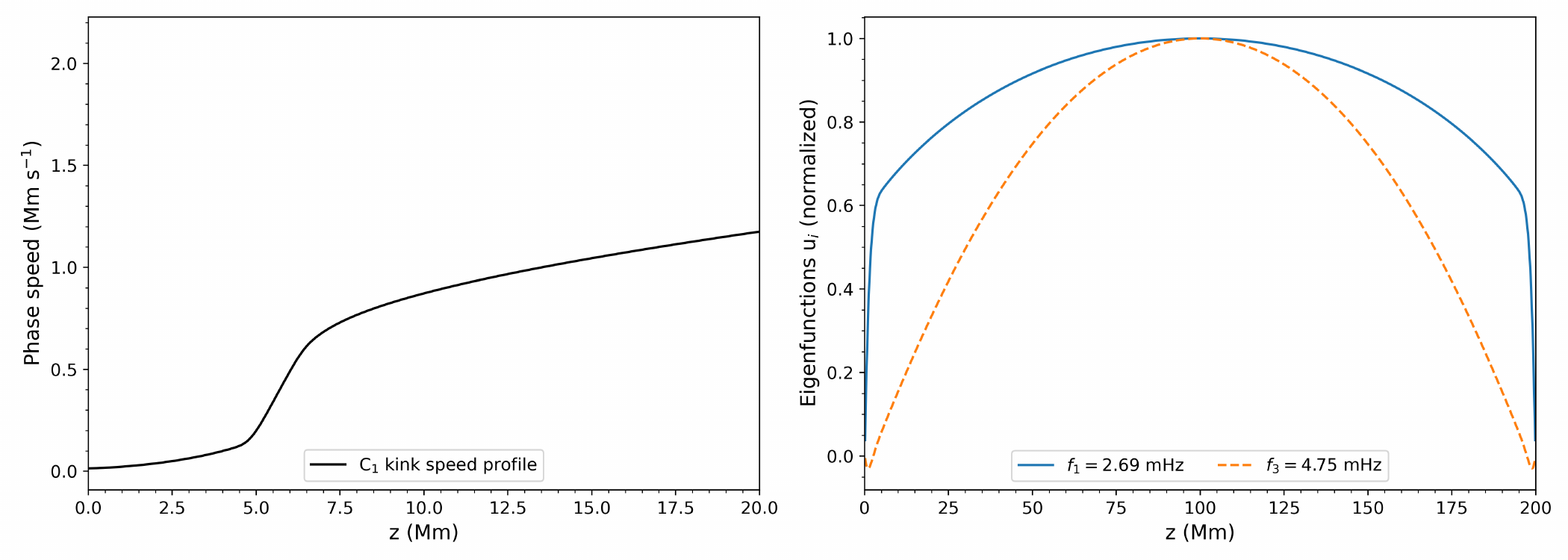}}
    \resizebox{\hsize}{!}{
    \includegraphics[trim={0.5cm 1.cm 0.4cm 0.3cm},clip,scale=0.25]{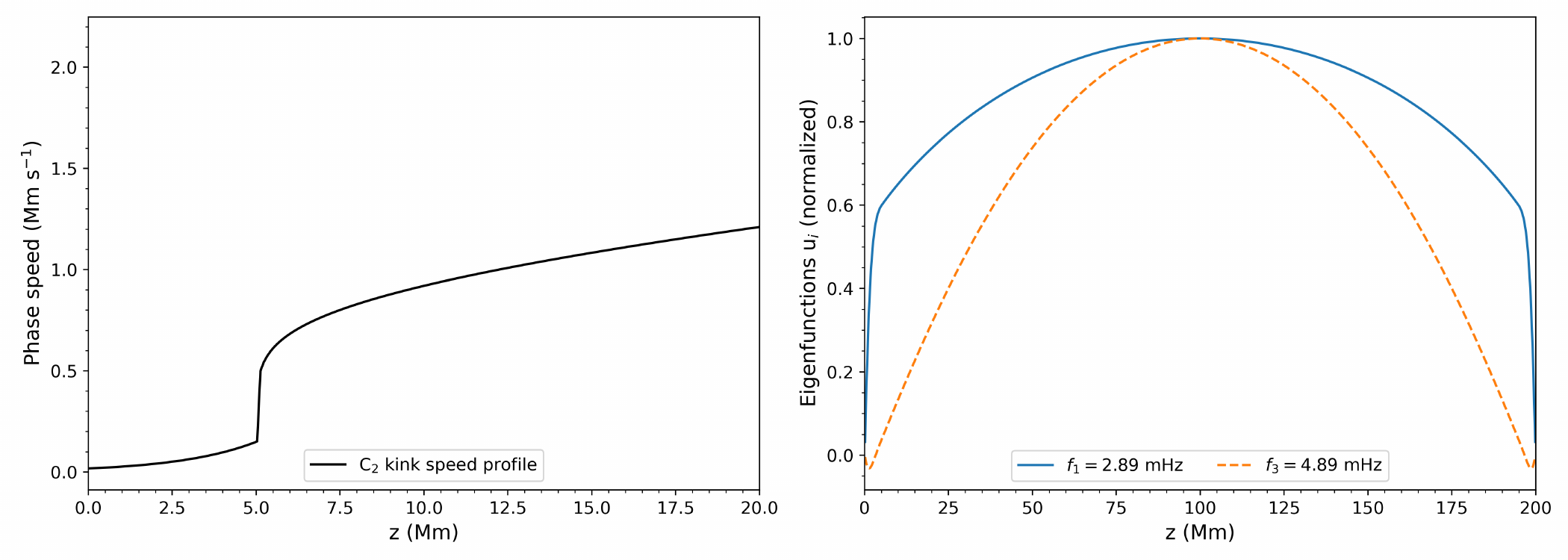}}
    \resizebox{\hsize}{!}{
    \includegraphics[trim={0.5cm 1.cm 0.4cm 0.3cm},clip,scale=0.25]{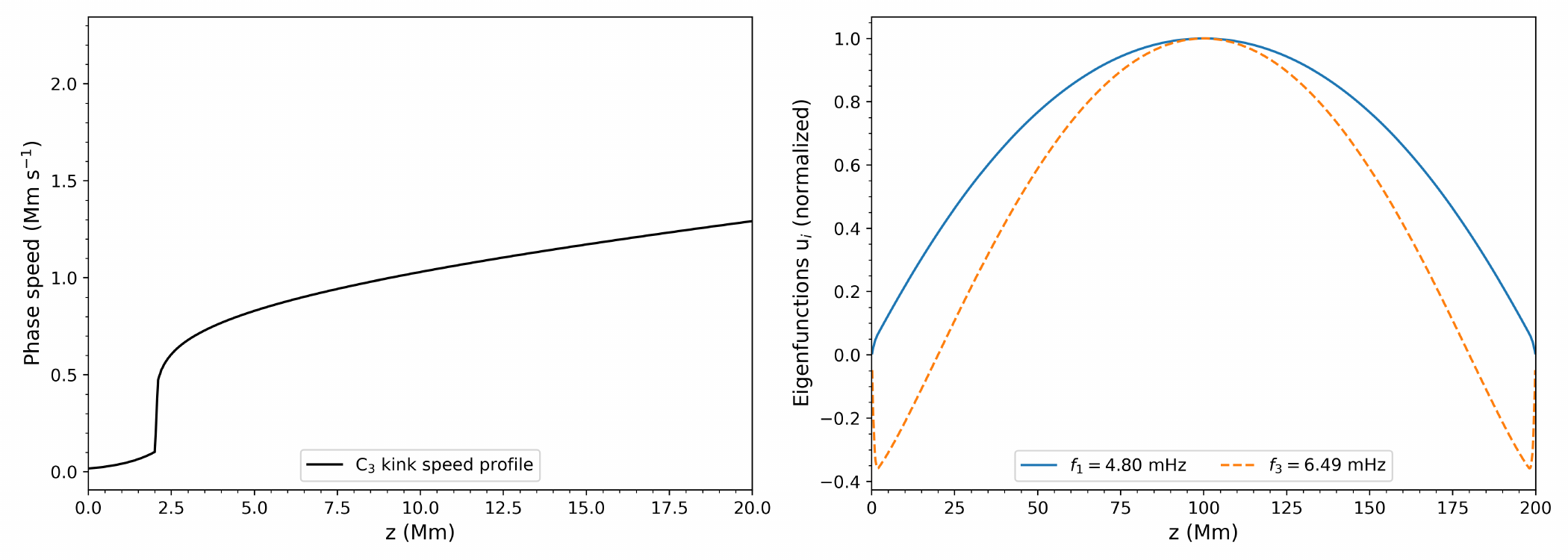}}
    \resizebox{\hsize}{!}{
    \includegraphics[trim={0.5cm 1.cm 0.4cm 0.3cm},clip,scale=0.25]{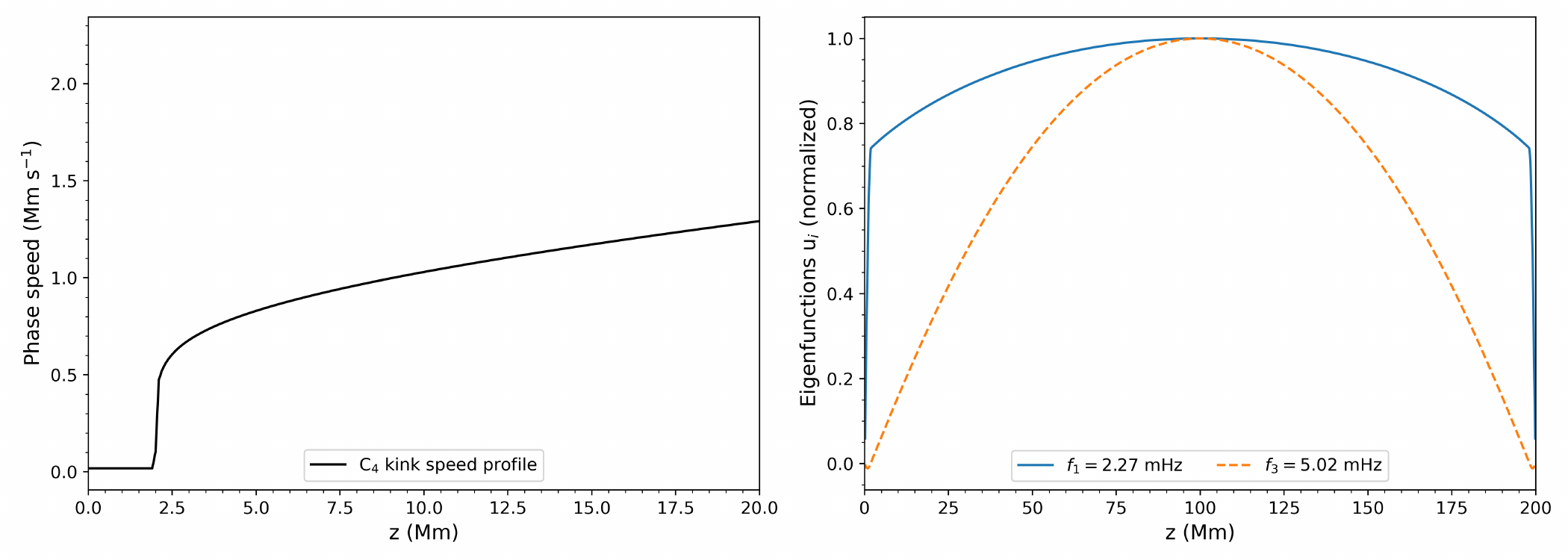}}
    \resizebox{\hsize}{!}{
    \includegraphics[trim={0.5cm 0.4cm 0.4cm 0.3cm},clip,scale=0.25]{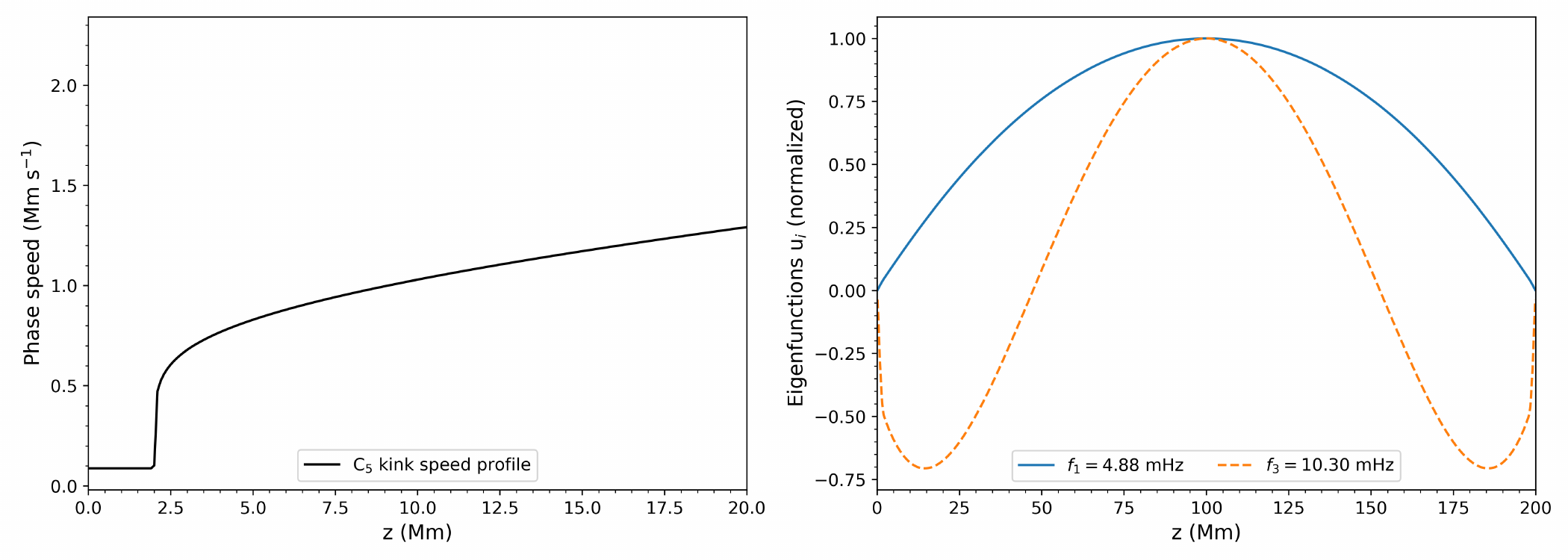}}   
    \caption{Different kink speed profiles used as phase speeds in Eq. \ref{eq:1D} and its respective solutions. Left: Kink profiles near one footpoint. Right: Corresponding first (solid blue lines) and third (dashed orange lines) harmonics  along the entire length of the loop. }\label{fig:param}
\end{figure}

In Sect. \ref{sec:1D} we solve the Sturm-Liouville problem of Eq. \ref{eq:1D} to find the modes and frequencies of non-axisymmetric, transverse oscillations of loops, considering the thin tube approximation. To further explore the effects that the profile of the lower solar atmosphere has on the solutions of Eq. \ref{eq:1D}, we performed a parameter study that considers different phase speed profiles. We used a set of five profiles ($C_1$ to $C_5$) that explores the effects of the width of the chromosphere and the transition region, as well as the values of the kink speed in the chromosphere with respect to the corona. The $C_1$ profile is the kink speed of our post-relaxation 2D loop profile, as seen in Fig. \ref{fig:inicon}. $C_2$ is the kink speed derived from the initial conditions of our pre-relaxation 2D profile, and is characterised by a sharper transition region. Both $C_1$ and $C_2$ are derived for a loop with a wide chromosphere ($\Delta z = 5$\,Mm). $C_3$ is a kink speed profile derived for a loop with a narrower chromosphere ($\Delta z = 2$\,Mm). $C_4$ and $C_5$ are modified versions of $C_3$, for uniform chromospheric values of the kink speed.

The results of our parameter study can be seen in Fig. \ref{fig:param}. Comparing the harmonics for the $C_1$ and $C_2$, we see that the width of the transition region, for the range considered here, does not generate qualitatively different results regarding the shape of the kink modes or the values of the frequencies for each harmonic. Reducing the width of the chromosphere ($C_3$) can influence the shape of the kink modes and their frequencies substantially, but the results are highly affected by the properties of the chromosphere and the shape of the resulting kink speed profile there ($C_4$, $C_5$). For example, having chromospheric kink speeds values that are less than $10\%$ of the coronal ones ($C_4$) can still lead to results similar to those presented for the $C_1$ profile, even for a narrow chromosphere with a sharp transition region.

We conclude that the width of the solar chromosphere and the local properties that define the kink speed of a loop are of great importance when deriving the kink modes of the reduced 1D Sturm-Liouville problem of Eq. \ref{eq:1D}.

\end{appendix}
\end{document}